\newif\ifshowrevisions
 \showrevisionsfalse   

\ifshowrevisions
    
    \newenvironment{revision}
        {\begingroup\color{black}}
        {\endgroup}
\else
    
    \newenvironment{revision}
        {}
        {}
\fi

\documentclass[5p,twocolumn,authoryear,times,nopreprintline]{elsarticle}
\usepackage{epsfig,amsmath,amssymb,natbib,color}
\usepackage{graphicx}
\usepackage{xcolor}
\usepackage{booktabs}
\usepackage{array}
\usepackage{tabularx}
\usepackage{makecell}
\usepackage[T1]{fontenc}
\usepackage[utf8]{inputenc}
\usepackage{xcolor}

\newcommand{\q}[1]{\textquotedblleft #1\textquotedblright}

\newcommand{\comm}[1]{}

\journal{International Journal of Multiphase Flow}
\usepackage{author-manuscript}

\begin{document}

\begin{frontmatter}

\title{Development and assessment of a target-$\omega$ interfacial turbulence-damping model for stratified gas--liquid flow\tnoteref{authorversion}}
\tnotetext[authorversion]{Author accepted manuscript. Published in \textit{International Journal of Multiphase Flow} 204 (2026), 105952. Version of record: \href{https://doi.org/10.1016/j.ijmultiphaseflow.2026.105952}{doi:10.1016/j.ijmultiphaseflow.2026.105952}. \copyright\ 2026. This manuscript version is made available under the \href{https://creativecommons.org/licenses/by-nc-nd/4.0/}{CC BY-NC-ND 4.0 license}.}

\author[label1]{Di Wu}

\author[label1]{Junda Liu}

\author[label1]{Qiaopeng Wang}

\author[label1]{Liu Liu}

\author[label1]{Dongling Wu\corref{cor1}}
\ead{dongling.wu@csu.edu.cn}

\author[label1]{Ping Zhou\corref{cor1}}
\ead{zhoup@csu.edu.cn}

\author[label2]{Pengyu Shi\corref{cor1}}
\ead{pengyu.shi@ulb.be}

\cortext[cor1]{Corresponding authors.}

\address[label1]{School of Energy Science and Engineering, Central South University, Lushan South Road No. 932, Changsha, 410083, China}

\address[label2]{Department of Transfers, Interfaces, and Processes, Universit\'e Libre de Bruxelles, 1050 Brussels, Belgium}

\begin{abstract}
Interfacial turbulence damping is required in Reynolds-averaged simulations of stratified gas--liquid flows when strong interfacial velocity gradients are present. In this work, we develop a target-$\omega$ interfacial damping formulation for the low-Reynolds-number Wilcox-type $k$--$\omega$ model in a one-field RANS/VOF framework. A gas-side target value of the specific dissipation rate $\omega$ is constructed from an effective interfacial wall coordinate determined from the local resolved flow. \textcolor{black}{The constructed target-$\omega$ provides a smooth transition between the viscous and logarithmic limits through power-law blending and requires no calibration of model constants against target data.} \textcolor{black}{The formulation is first assessed against three gas-layer DNS cases at $Re_\tau\simeq180$ with different interface velocities. The model reproduces the principal features of the mean velocity, turbulent kinetic energy and Reynolds shear-stress distributions, while showing substantially weaker sensitivity to the near-interface grid resolution than the Egorov-type formulation, although some grid dependence remains on the coarsest mesh.} The assessment is then extended to the co-current Fabre stratified-flow experiment at a higher Reynolds number \textcolor{black}{and to the counter-current St\"abler experiment}. \textcolor{black}{In both cases, the present formulation provides generally good predictions of the gas-side mean flow and turbulence statistics without case-specific adjustment. For the St\"abler case, in which the liquid level develops as part of the coupled two-phase solution, the predicted mean liquid depth agrees closely with the measured value.} \textcolor{black}{These results indicate that the target-$\omega$ formulation provides a consistent interfacial turbulence-damping treatment for resolved large gas--liquid interfaces over the range of conditions considered.}
\end{abstract}

\begin{keyword}
Interfacial turbulence damping \sep Stratified gas--liquid flow \sep RANS/VOF \sep $k$--$\omega$ turbulence model \sep Target-$\omega$ formulation
\end{keyword}

\end{frontmatter}


\section{Introduction}
\label{sec:introduction}

Gas--liquid two-phase flows in channels and pipelines exhibit a variety of flow patterns, including stratified, stratified-wavy and plug flows \citep{mahdi2024}. Among these, stratified flow is distinguished by the presence of a persistent large-scale gas--liquid interface separating the two phases. In its smooth-interface form, this regime is favoured when interfacial disturbances are suppressed, for example by large density and viscosity contrasts and sufficiently high interfacial tension \citep{aldogail2021}. Such configurations are encountered in applications such as gas--oil pipeline transport and gas-heated molten-metal flows \citep{chi2019,yan2024}, where the momentum exchange across the interface strongly affects the mean flow and the near-interface turbulence.

For industrial-scale simulations of stratified flows, resolving all turbulent scales is generally impractical, and modelling based on the Reynolds-averaged Navier--Stokes (RANS) equations remains one of the most practical frameworks for engineering calculations involving large gas--liquid interfaces. For example, \citet{frederix2018} used a two-fluid RANS framework to analyse turbulence damping near large-scale interfaces, while \citet{burdairon2023} used a combined VOF--RANS approach to study wind-driven interfacial waves. In such calculations, the effect of unresolved turbulent fluctuations on the mean flow is represented through turbulence closures, often based on the eddy-viscosity hypothesis and implemented through one- or two-equation models.

However, modelling turbulence in the immediate vicinity of a gas--liquid interface requires particular care. The local behaviour of the Reynolds stresses and of the turbulence-model variables must be consistent with the interfacial constraints. In large-interface RANS simulations, standard differential eddy-viscosity models, such as the $k$--$\varepsilon$ and $k$--$\omega$ models, may generate non-physical turbulence levels near the interface when large interfacial velocity gradients are present. This issue has motivated the development of interfacial turbulence-damping treatments, including the wall-like source-term formulation of \citet{egorov2004}, the grid-related reformulation of \citet{frederix2018}, and the density-based two-fluid damping model of \citet{dong2022}. A summary of representative formulations is provided in \ref{app:damping_models}.

The present study uses the low-Reynolds-number Wilcox $k$--$\omega$ closure \citep{wilcox1988} as the underlying RANS model. Within this closure, the turbulent viscosity is related to $k/\omega$, so that an increase of $\omega$ near the interface reduces the local turbulent viscosity. The interfacial damping treatments considered here therefore act primarily through the equation for the specific dissipation rate $\omega$, either by adding a source term $S_\omega$ or by imposing an equivalent near-interface prescription. In particular, many damping formulations are designed to increase the local value of $\omega$ in the interfacial region, thereby suppressing the excessive turbulent viscosity generated by large interfacial velocity gradients. This additional interfacial contribution is absent from the baseline $k$--$\omega$ closure. The complete governing equations and implementation details are given in \S~\ref{sec:model}.

Only the aspects of existing damping treatments that are directly relevant to the present formulation are recalled here. A relatively direct approach is to treat a smooth interface as an additional wall-like damping surface in low-Reynolds-number $k$--$\varepsilon$ models, using the shortest distance to either the wall or the interface in the near-boundary damping functions, as in the treatments of \citet{berthelsen2007} and \citet{duan2014}. Other approaches prescribe an interfacial condition for the specific dissipation rate, as in the free-surface treatment of \citet{kamath2019}. A widely used alternative is to introduce explicit source terms into the turbulence transport equations. Among these, the model of \citet{egorov2004} is particularly influential because it can be implemented by adding a wall-like source term to the $\omega$ equation of a $k$--$\omega$-type closure.

For phase $i$, the Egorov source term may be written as
\begin{equation}
\label{eq:egorov_intro}
S^E_{\omega,i}
=
A_i\Delta_\perp\beta\rho_i
\left(
B\frac{6\mu_i}{\beta\rho_i\Delta_\perp^2}
\right)^2,
\qquad i=G,L,
\end{equation}
where $G$ and $L$ denote the gas and liquid phases, respectively. Here, $A_i$ is the interfacial area density, $\Delta_\perp$ denotes the interface-normal cell-thickness measure used to represent the typical cell size across the interface, $\rho_i$ and $\mu_i$ are the density and dynamic viscosity of phase $i$, respectively, and $B$ is an empirical damping factor. The factor $A_i$ activates the source term near the interface. By increasing the local value of $\omega$, the source term reduces the turbulent viscosity through the usual $k$--$\omega$ relation between $\mu_t$ and $k/\omega$. Owing to its simple implementation, the Egorov-type model has subsequently been adopted in a number of large-interface, free-surface and annular-flow simulations, for example by \citet{porombka2015}, \citet{gada2017}, \citet{wen2018} and \citet{fan2019}.

Several modifications or related formulations have been proposed following this idea. \citet{fan2019} addressed two limitations of the original Egorov model: its dependence on mesh size and its symmetric treatment of the two phases. They introduced an interfacial length scale less sensitive to tangential mesh dimensions and proposed an asymmetric source-term formulation. \citet{varallo2024} introduced an analogous Egorov-type correction into the $\varepsilon$ equation of an RNG $k$--$\varepsilon$ model for annular-flow simulations. A different route was taken by \citet{dong2022}, who proposed a density-based damping model within a two-fluid RANS framework by analysing the per-phase turbulence transport equations. Their formulation reduces the need for empirical calibration in that framework and introduces an asymmetric damping mechanism based on the local phase distribution and density contrast. In contrast, the present work considers interfacial damping within a one-field RANS/VOF framework using a $k$--$\omega$ closure, for which the calibration and grid-dependence issues associated with the empirical prefactor $B$ in Egorov-type treatments remain unresolved. 

Equation~\eqref{eq:egorov_intro} also makes clear why the damping strength is difficult to prescribe in a universal manner. The imposed near-interface value of $\omega$ scales with $B/\Delta_\perp^2$, so that the same value of $B$ does not necessarily correspond to the same physical damping strength on different grids. In the original report, \citet{egorov2004} stated that $B$ should be sufficiently large, at least 10, in the corresponding wall-boundary-condition formulation, and used $B=100$ in their free-surface source-term implementation because the interfacial grids were relatively coarse. Subsequent applications of Egorov-type or related damping formulations have adopted a wide range of values. For example, \citet{zhang2025} used $B=10$ in the Fan-type complex damping formulation for inclined annular-flow simulations; \citet{wen2018} tested $B=0$, $15$ and $30$ in condensing mini/macro-channel simulations and selected $B=15$ by comparison with heat-transfer correlations; \citet{fan2019} selected $B=25$ based on annular-flow wave data; \citet{porombka2015} used values around $B=100$; and \citet{gada2017} adopted $B=500$ in their large-scale-interface model. \citet{frederix2018} further showed that the effective damping produced by the Egorov formulation depends strongly on the interfacial grid size. Taken together, the wide range of reported values of $B$ and this grid-size sensitivity indicate that the effective damping strength depends on the flow configuration, grid resolution and calibration criterion. A generally applicable prescription for $B$ is therefore still lacking.

Motivated by these limitations, the present study develops a target-$\omega$ interfacial damping formulation using the Wilcox-type $k$--$\omega$ model as the underlying RANS closure. Rather than prescribing the interfacial damping strength through an empirical factor such as $B$, the target value of $\omega$ is determined from an effective interfacial wall coordinate obtained from the local resolved flow. \textcolor{black}{The resulting target combines the viscous and logarithmic near-wall limits and is applied through a gas-side interfacial source term. No constant in the target-$\omega$ relation is fitted to the DNS or experimental data considered in the present study. The aim is therefore to provide a physically constrained control of the gas-side turbulent viscosity that requires no case-specific fitting of an interfacial damping coefficient and reduces the explicit sensitivity of the damping treatment to the near-interface grid resolution.}

The remainder of the paper is organised as follows. Section~\ref{sec:model} presents the target-$\omega$ damping formulation and its implementation in the $k$--$\omega$ framework. Section~\ref{sec:configuration} introduces the canonical smooth-interface configuration, the DNS reference data and the corresponding RANS/VOF calculations. \textcolor{black}{Section~\ref{sec:results} presents the DNS-based validation in the canonical configuration and the subsequent assessments against the stratified-flow experiments of \citet{fabre1987experimental} and \citet{staebler2007}.} Finally, Section~\ref{sec:conclusions} summarises the main findings and outlines possible extensions of the model.

\section{Target-\texorpdfstring{$\omega$}{omega} interfacial damping formulation}
\label{sec:model}

The present formulation is developed for a smooth stratified gas--liquid interface. In the configurations considered here, the liquid phase remains laminar or weakly turbulent, so that the interfacial turbulence correction is intended to damp the gas-side turbulence near the interface. The turbulence closure is solved in the one-field RANS/VOF framework used in the present calculations, where a single $k$ field and a single $\omega$ field are transported with mixture properties. With the interfacial correction included as an additional source term, the transport equation for the specific dissipation rate reads
\begin{equation}
\label{eq:omega_transport_model}
\rho\frac{\mathrm{D}\omega}{\mathrm{D}t}
=
\alpha\frac{\omega}{k}P_k
-\beta\rho\omega^2
+\frac{\partial}{\partial x_j}
\left[
\left(\mu+\sigma_\omega\mu_t\right)
\frac{\partial\omega}{\partial x_j}
\right]
+S_{\omega}^{I}.
\end{equation}
Here, $k$, $\omega$, $\rho$, $\mu$ and $\mu_t$ are the one-field turbulent kinetic energy, specific dissipation rate, density, dynamic viscosity and turbulent dynamic viscosity, respectively. The term $P_k$ denotes the production of turbulent kinetic energy, and $\alpha$, $\beta$ and $\sigma_\omega$ are model constants. The additional term $S_{\omega}^{I}$ denotes the interfacial damping source. Its target value is constructed from gas-side quantities, while the source itself is localised on the gas side of the reconstructed interface. No additional source term is introduced in the $k$ equation in the present formulation; the interfacial correction modifies the turbulent viscosity indirectly through its effect on $\omega$. 

The interfacial correction is introduced as a positive source term in the $\omega$ equation, following the same general source-term structure as Egorov-type damping treatments. In the baseline $k$--$\omega$ equation, $\omega$ is locally reduced by the destruction term $-\beta\rho\omega^2$. The additional source is therefore constructed from a prescribed target value $\omega_{\mathrm{tar}}$ and a local gas-side activation factor. The gas-side source is written as
\begin{equation}
\label{eq:source_target_omega}
S_{\omega}^{I}
=
A_g\Delta_\perp\beta\rho\omega_{\mathrm{tar}}^2 .
\end{equation}
Here, $\rho$ is the local one-field density and $\Delta_\perp$ is the local mesh thickness in the direction normal to the reconstructed interface. \textcolor{black}{The source is localised using a binary pure-gas indicator $\chi_g$ defined from the VOF gas volume fraction $\alpha_g$ as}
\begin{equation}
\label{eq:pure_gas_indicator}
{\color{black}
\chi_g=
\begin{cases}
1, & \alpha_g>0.999,\\
0, & \alpha_g\le0.999.
\end{cases}
}
\end{equation}
\textcolor{black}{The corresponding gas-side interfacial area density is defined as}
\begin{equation}
\label{eq:gas_side_area_density}
{\color{black}
A_g=2\chi_g|\nabla\chi_g|.
}
\end{equation}

\begin{figure*}[!t]
    \centering
    \includegraphics[width=1.0\linewidth]{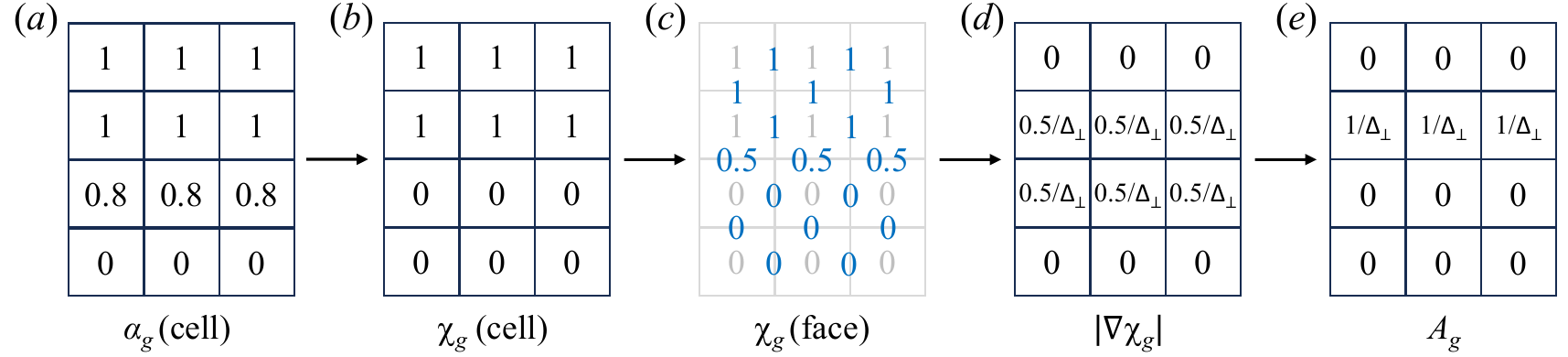}
    \caption{\textcolor{black}{Schematic illustration of the construction of the gas-side interfacial area density for an idealised flat interface aligned with a uniform mesh. ($a$) Cell-centred VOF gas volume fraction $\alpha_g$, including a layer of mixed cells representative of the interfacial region. ($b$) Cell-centred pure-gas indicator $\chi_g$, obtained by setting $\chi_g=1$ for $\alpha_g>0.999$ and $\chi_g=0$ otherwise. ($c$) Face values of $\chi_g$ obtained by interpolation. ($d$) Magnitude of the reconstructed cell-centred gradient $|\nabla\chi_g|$; the cells immediately adjacent to the transition in $\chi_g$ have $|\nabla\chi_g|=0.5/\Delta_\perp$ in this idealised configuration. ($e$) Gas-side interfacial area density $A_g=2\chi_g|\nabla\chi_g|$. Multiplication by $2\chi_g$ suppresses the contribution on the mixed/liquid side, giving $A_g=1/\Delta_\perp$ in the first pure-gas cell layer adjacent to the interfacial region and zero elsewhere.}}
    \label{fig:Ag-localisation}
\end{figure*}

\textcolor{black}{Figure~\ref{fig:Ag-localisation} illustrates the construction of $A_g$ for an idealised flat interface aligned with a uniform mesh. Figure~\ref{fig:Ag-localisation}($a$) shows the original cell-centred VOF gas volume fraction, including a layer of mixed cells representative of the interfacial region. Applying the pure-gas criterion in Eq.~\eqref{eq:pure_gas_indicator} gives the binary indicator $\chi_g$ shown in figure~\ref{fig:Ag-localisation}($b$), which is subsequently interpolated to the cell faces [figure~\ref{fig:Ag-localisation}($c$)]. The resulting cell-centred gradient magnitude is shown in figure~\ref{fig:Ag-localisation}($d$). Multiplication by $2\chi_g$ suppresses the contribution on the mixed/liquid side and gives the gas-side interfacial area density shown in figure~\ref{fig:Ag-localisation}($e$). For this limiting mesh-aligned configuration, $A_g=1/\Delta_\perp$ in the first pure-gas cell layer adjacent to the interfacial mixed-cell region and is zero elsewhere, so that $A_g\Delta_\perp=1$ in the activated cells. The binary field $\chi_g$ is used solely to localise the gas-side damping source, whereas geometric quantities such as the wall-law distance $d_P$ are evaluated from the reconstructed interface. The numerical implementation of this interface representation is described in \S~\ref{sec:rans_vof}. }

The key difference from the original Egorov formulation lies in the construction of the near-interface target value of $\omega$. \citet{egorov2004} based their formulation on the standard near-wall behaviour of $k$--$\omega$ models, for which the specific dissipation rate scales as $6\nu/(\beta y^2)$ close to a wall \citep{wilcox1988}. In some practical $k$--$\omega$ implementations, this wall value is multiplied by an empirical factor to improve robustness \citep{menter1994}. The Egorov-type interfacial prescription can therefore be written as
\begin{equation}
\label{eq:egorov_target_omega}
\omega_{\mathrm{E}}
=
B\frac{6\mu_g}{\beta\rho_g\Delta_\perp^2},
\end{equation}
where $B$ is an empirical damping factor. In contrast, the present formulation does not prescribe the interfacial value of $\omega$ through an empirical damping factor. Instead, a target value $\omega_{\mathrm{tar}}$ is determined from an effective interfacial wall coordinate obtained from the local resolved flow through the wall-law relation described below.

To construct $\omega_{\mathrm{tar}}$, we introduce an effective interfacial friction velocity, $u_{\tau}^{\mathrm{eff}}$. \textcolor{black}{Rather than evaluating the interfacial shear stress directly, $u_{\tau}^{\mathrm{eff}}$ is obtained from a local wall-law inversion based on the resolved velocity variation between the reconstructed interface and the first pure-gas cell.} Based on $u_{\tau}^{\mathrm{eff}}$, the characteristic scales for velocity, length and specific dissipation rate are chosen as
\begin{equation}
\label{eq:wall_unit_scales}
u_{\tau}^{\mathrm{eff}},\qquad
\frac{\mu_g}{\rho_g u_{\tau}^{\mathrm{eff}}},\qquad
\frac{\rho_g\left(u_{\tau}^{\mathrm{eff}}\right)^2}{\mu_g},
\end{equation}
respectively. Unless otherwise stated, all quantities carrying the superscript $\ast$ in this section are dimensionless quantities normalised using the corresponding scales in \eqref{eq:wall_unit_scales}. In particular, a normal distance $y$ from the reconstructed interface is expressed in these effective wall units as $y^\ast$.

The viscous-sublayer estimate is adopted from the correlation-based $y^+$-insensitive near-wall treatment for $\omega$-based turbulence models in ANSYS Fluent \citep{ansys2024fluent},
\begin{equation}
\label{eq:omega_vis_star}
{
\omega_{\mathrm{vis}}^\ast
=
\textcolor{black}{C_F}\frac{6}{\beta\left(y^\ast\right)^2},
}
\end{equation}
\textcolor{black}{where $C_F=1/3$.} In the logarithmic limit, the corresponding equilibrium $k$--$\omega$ estimate is
\begin{equation}
\label{eq:omega_log_star}
\omega_{\mathrm{log}}^\ast
=
\frac{1}{C_\mu^{1/2}\kappa y^\ast}.
\end{equation}
\textcolor{black}{The two asymptotic estimates are combined through the power-law blending relation}
\begin{equation}
\label{eq:omega_target_star}
{\color{black}
\omega_{\mathrm{tar}}^\ast
=
\left[
\left(\omega_{\mathrm{vis}}^\ast\right)^p
+
\left(\omega_{\mathrm{log}}^\ast\right)^p
\right]^{1/p},
}
\end{equation}
\textcolor{black}{with}
\begin{equation}
\label{eq:fixed_target_constants}
{\color{black}
p=1.3,\qquad
C_\mu=0.09,\qquad
\kappa=0.4187.
}
\end{equation}
\textcolor{black}{The coefficient $C_F$ and the blending exponent $p=1.3$ are adopted from the correlation-based near-wall treatment described in Section~4.18.3.2 of the ANSYS Fluent Theory Guide~\citep{ansys2024fluent}. All constants entering the target-$\omega$ relation are fixed \emph{a priori} and are not fitted to the DNS or experimental data considered in the present study.}

The dimensional target value is obtained as
\begin{equation}
\label{eq:omega_target_dimensional}
\omega_{\mathrm{tar}}
=
\omega_{\mathrm{tar}}^\ast
\frac{\rho_g\left(u_{\tau}^{\mathrm{eff}}\right)^2}{\mu_g}.
\end{equation}

\textcolor{black}{The remaining task is to determine $u_\tau^{\mathrm{eff}}$. The iterative procedure follows the wall-law inversion concept of \citet{knopp2006}, but the matching relation is reformulated here as a scalar nonlinear equation for the effective wall coordinate. For the first pure-gas cell adjacent to the interfacial region, let $d_P$ denote the normal distance from the cell centre to the reconstructed interface. The local velocity difference entering the wall-law relation is approximated as}
\begin{equation}
\label{eq:delta_u_gradient}
{\color{black}
\Delta u = G_P d_P,
}
\end{equation}
\textcolor{black}{where $G_P$ is the magnitude of the resolved velocity-gradient tensor at the cell centre,}
\begin{equation}
\label{eq:velocity_gradient_magnitude}
{\color{black}
G_P
=
\left(
\frac{\partial u_i}{\partial x_j}
\frac{\partial u_i}{\partial x_j}
\right)^{1/2}.
}
\end{equation}
\textcolor{black}{For the canonical parallel-flow configurations considered below, $G_P$ reduces to the magnitude of the tangential-velocity gradient normal to the interface, $|\partial u_t/\partial n|$. Because $G_P$ is a magnitude, the velocity scale entering the wall-law inversion remains non-negative under local flow reversal.}

\textcolor{black}{The iteration is initialised with the local stress-based estimate}
\begin{equation}
\label{eq:initial_friction_velocity}
{\color{black}
u_\tau^{(0)}
=
\left[
\frac{
\mu_g G_P
+
C_\mu^{1/2}\rho_g k
}
{\rho_g}
\right]^{1/2}.
}
\end{equation}
\textcolor{black}{This gives the initial effective wall coordinate $d_{P,0}^\ast=\rho_g u_\tau^{(0)}d_P/\mu_g$. The Reichardt-type wall-law matching relation \citep{reichardt1951,pope2000} is}
\begin{equation}
\label{eq:reichardt_wall_law}
\begin{aligned}
\Delta u^\ast
&=\frac{1}{\kappa}\ln\left(1+\kappa d_P^\ast\right)\\
&\quad+7.8\left[1-\exp\left(-\frac{d_P^\ast}{11}\right)
-\frac{d_P^\ast}{11}\exp\left(-\frac{d_P^\ast}{3}\right)\right].
\end{aligned}
\end{equation}

\textcolor{black}{Using $\Delta u=G_Pd_P$, together with}
\begin{equation}
\label{eq:wall_units_for_iteration}
{\color{black}
\Delta u^\ast=\frac{\Delta u}{u_\tau^{\mathrm{eff}}},
\qquad
d_P^\ast=\frac{\rho_g u_\tau^{\mathrm{eff}}d_P}{\mu_g},
}
\end{equation}
\textcolor{black}{the wall-law matching condition can be recast as}
\begin{equation}
\label{eq:wall_law_scalar}
{\color{black}
\mathcal{G}\left(d_P^\ast\right)
=
\frac{\rho_gG_Pd_P^2}{\mu_gd_P^\ast}
-
F_R\left(d_P^\ast\right)
=
0,
}
\end{equation}
\textcolor{black}{where $F_R$ denotes the right-hand side of Eq.~\eqref{eq:reichardt_wall_law}. Equation~\eqref{eq:wall_law_scalar} is solved directly for $d_P^\ast$ using an under-relaxed Newton iteration,}
\begin{equation}
\label{eq:newton_update}
{\color{black}
d_{P,n+1}^\ast
=
d_{P,n}^\ast
-
0.5
\frac{
\mathcal{G}\left(d_{P,n}^\ast\right)
}{
\mathcal{G}'\left(d_{P,n}^\ast\right)
}.
}
\end{equation}
\textcolor{black}{A maximum of ten Newton iterations is allowed. During the iteration, $d_P^\ast$ is bounded below by $10^{-12}$ to prevent non-physical negative values. Let $\delta_N=\mathcal{G}/\mathcal{G}'$ denote the unrelaxed Newton correction. Convergence is accepted when $|\delta_N/d_P^\ast|<10^{-10}$ after the update. If this criterion is not satisfied within ten iterations, the final iterate is retained for the current source-term evaluation and the local wall-law inversion is repeated during the next global nonlinear solver iteration.}

\textcolor{black}{After convergence, the effective friction velocity is recovered from}
\begin{equation}
\label{eq:effective_friction_velocity}
{\color{black}
u_\tau^{\mathrm{eff}}
=
\frac{\mu_gd_P^\ast}{\rho_gd_P}.
}
\end{equation}
\textcolor{black}{The effective wall coordinate entering Eqs.~\eqref{eq:omega_vis_star}--\eqref{eq:omega_target_star} is then set to $y^\ast=d_P^\ast$. The dimensional target $\omega_{\mathrm{tar}}$ is evaluated from Eq.~\eqref{eq:omega_target_dimensional}, after which the gas-side interfacial source is obtained from Eq.~\eqref{eq:source_target_omega}.}

\section{Computational configurations and validation strategy}
\label{sec:configuration}

The validation strategy is organised as follows. We first define a canonical smooth-interface channel configuration in which interfacial waves and interface deformation are deliberately excluded. This configuration provides a controlled setting for examining the near-interface behaviour of interfacial turbulence-damping models. We then describe the gas-layer DNS calculations used to provide reference turbulence statistics for the smooth-interface gas flow. Finally, the corresponding two-phase RANS/VOF configuration is introduced, together with the numerical implementation of the proposed interfacial turbulence-damping model and the cases used to assess its performance over different prescribed interfacial velocities and near-interface grid resolutions.

\subsection{Canonical smooth-interface channel configuration}
\label{sec:canonical_configuration}

\begin{figure}[!tbp]
\centering
\includegraphics[width=\linewidth]{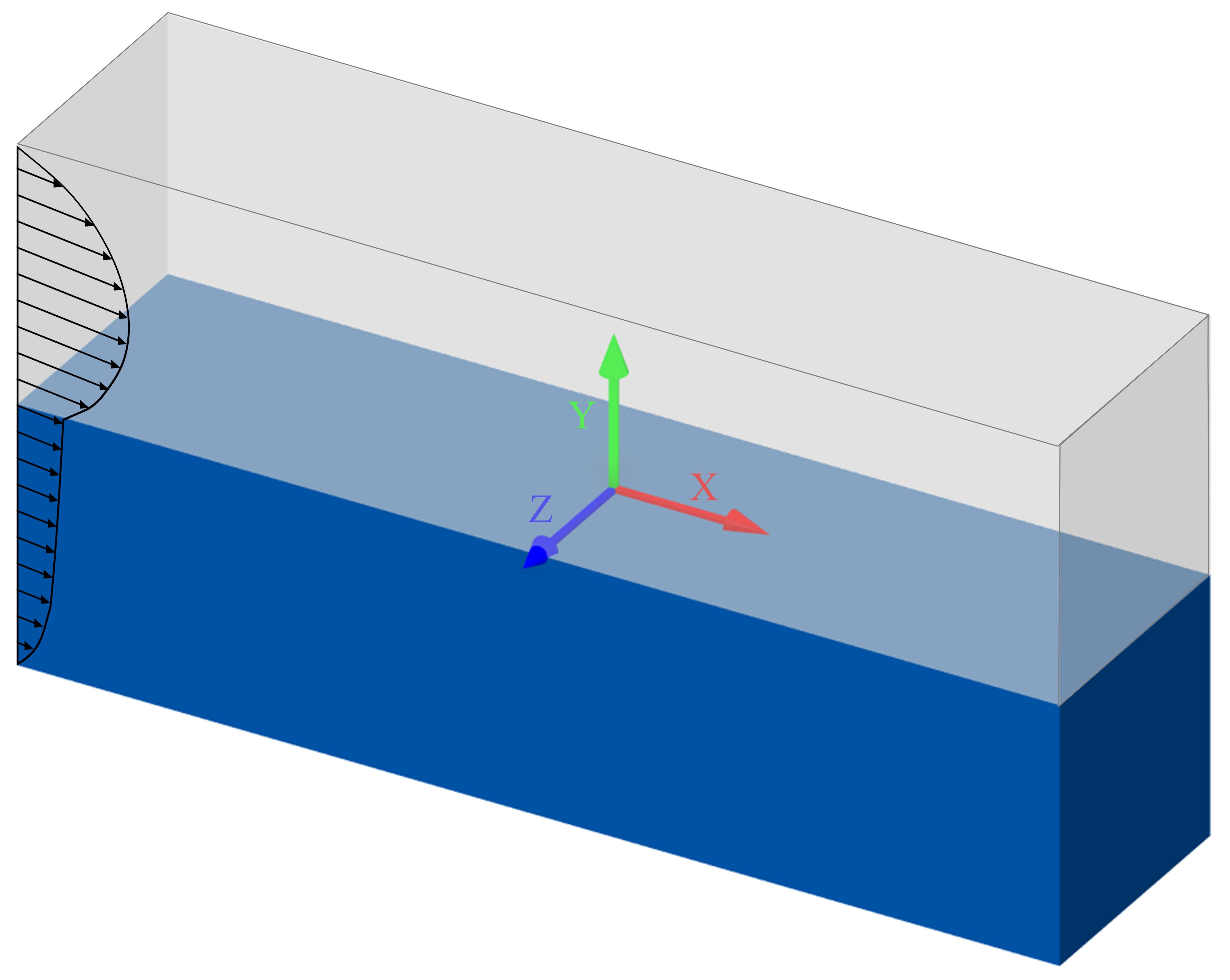}
\caption{Schematic of the canonical smooth-interface channel configuration. The gas--liquid interface is located at $y=0$, with the gas layer occupying $0<y<H$ and the liquid layer occupying $-H<y<0$.}
\label{fig:problem_setup}
\end{figure}

The model is assessed in a canonical smooth-interface channel configuration designed to isolate the effect of interfacial turbulence damping. As illustrated in figure~\ref{fig:problem_setup}, the flow consists of a horizontal gas layer over a liquid layer, separated by a flat gas--liquid interface. The configuration is designed to maintain a nearly smooth and horizontal interface, so that interfacial waves, entrainment and significant interface deformation are suppressed. This allows the comparison to focus on the behaviour of the turbulence closure in the immediate vicinity of a large, smooth interface.

A Cartesian coordinate system is used, with $x$ denoting the streamwise direction, $y$ the wall-normal direction and $z$ the spanwise direction. The gas--liquid interface is located at $y=0$, with the gas occupying $0<y<H_g$ and the liquid occupying $-H_l<y<0$. Throughout this work, we set $H_g=H_l=H$, so that $H$ denotes the height of each phase layer. In the gas layer, the normalised coordinate $y/H$ therefore increases from 0 at the gas--liquid interface to 1 at the upper wall. The streamwise and spanwise dimensions are chosen as $L_x=4H$ and $L_z=H$, respectively. Periodic boundary conditions are applied in the streamwise and spanwise directions, so that the computational domain represents a spatially periodic section of a fully developed stratified flow. The upper gas-side boundary and the lower liquid-side boundary are treated as no-slip, no-penetration walls.

The flow is driven by a constant streamwise pressure gradient, $G=-\mathrm{d}p/\mathrm{d}x>0$. This pressure gradient defines the gas-side pressure-gradient friction velocity
\begin{equation}
\label{eq:pressure_gradient_friction_velocity}
u_{\tau,p}
=
\left(
\frac{GH}{2\rho_g}
\right)^{1/2}.
\end{equation}
Based on $u_{\tau,p}$, the characteristic scales for velocity, length, turbulent kinetic energy, specific dissipation rate and dynamic turbulent viscosity are chosen as
\begin{equation}
\label{eq:pressure_gradient_scales}
u_{\tau,p},\qquad
\frac{\mu_g}{\rho_g u_{\tau,p}},\qquad
u_{\tau,p}^2,\qquad
\frac{\rho_g u_{\tau,p}^2}{\mu_g},\qquad
\mu_g,
\end{equation}
respectively. Unless otherwise stated, all quantities carrying the superscript $+$ in this section are dimensionless quantities normalised using the corresponding scales in \eqref{eq:pressure_gradient_scales}. In particular, a normal distance $y$ from the gas--liquid interface is expressed in pressure-gradient wall units as $y^+$.

The corresponding gas-side friction Reynolds number is defined as
\begin{equation}
\label{eq:friction_reynolds_number}
Re_\tau
=
\frac{\rho_g u_{\tau,p}(H/2)}{\mu_g},
\end{equation}
and the interfacial velocity is characterised by
\begin{equation}
\label{eq:interface_velocity_plus}
u_I^+
=
\frac{u_I}{u_{\tau,p}}.
\end{equation}
The value of $Re_\tau$ is determined by the gas properties, the gas-layer height and the imposed pressure gradient, while $u_I^+$ additionally depends on the prescribed interfacial velocity. Since a long-term motivation of this work is the modelling of iron-ore sintering processes, the material properties are chosen to be representative of air and a liquid-metal phase \citep{machida2005experimental}. Specifically, we set $\rho_g=1.225~\mathrm{kg\,m^{-3}}$, $\mu_g=1.789\times10^{-5}~\mathrm{Pa\,s}$, $\rho_l=3500~\mathrm{kg\,m^{-3}}$ and $\mu_l=0.05~\mathrm{Pa\,s}$. The phase-layer height is $H=5~\mathrm{mm}$ and the imposed pressure gradient is $G=542~\mathrm{Pa\,m^{-1}}$. With these specifications, the gas-side friction Reynolds number is fixed at $Re_\tau\simeq180$. Three interfacial velocities are considered, namely $u_I^+=0$, $0.95$ and $1.9$, corresponding to $u_I=0$, $1$ and $2~\mathrm{m\,s^{-1}}$, respectively. The way in which the prescribed interfacial velocity is imposed differs between the DNS and RANS calculations, and is described in the following two subsections.

The present configuration is not intended to reproduce the full dynamics of a wavy stratified flow. Instead, it provides a controlled smooth-interface benchmark in which the sensitivity of interfacial turbulence-damping models to the prescribed interfacial velocity and to the near-interface grid resolution can be examined systematically.

\subsection{Reference gas-layer DNS}
\label{sec:reference_dns}

Because the interface is constrained to remain flat and to move with a prescribed velocity, the reference DNS is performed only in the gas layer. The gas--liquid interface is therefore represented as a no-slip, no-penetration wall moving at the prescribed velocity $u_I$. The computational domain is a rectangular box of size $L_x\times L_y\times L_z=4H\times H\times H$. The flow is driven by the pressure gradient specified in \S~\ref{sec:canonical_configuration}. This DNS configuration is consistent with the simplified smooth-interface setup described above and provides the gas-side reference data used to assess the RANS damping models. The DNS therefore represents a limiting gas-side smooth-interface problem with a prescribed interfacial velocity, rather than a fully resolved two-phase DNS with interfacial deformation and two-sided stress coupling.

\begin{figure}[!tbp]
\centering
\includegraphics[width=\linewidth]{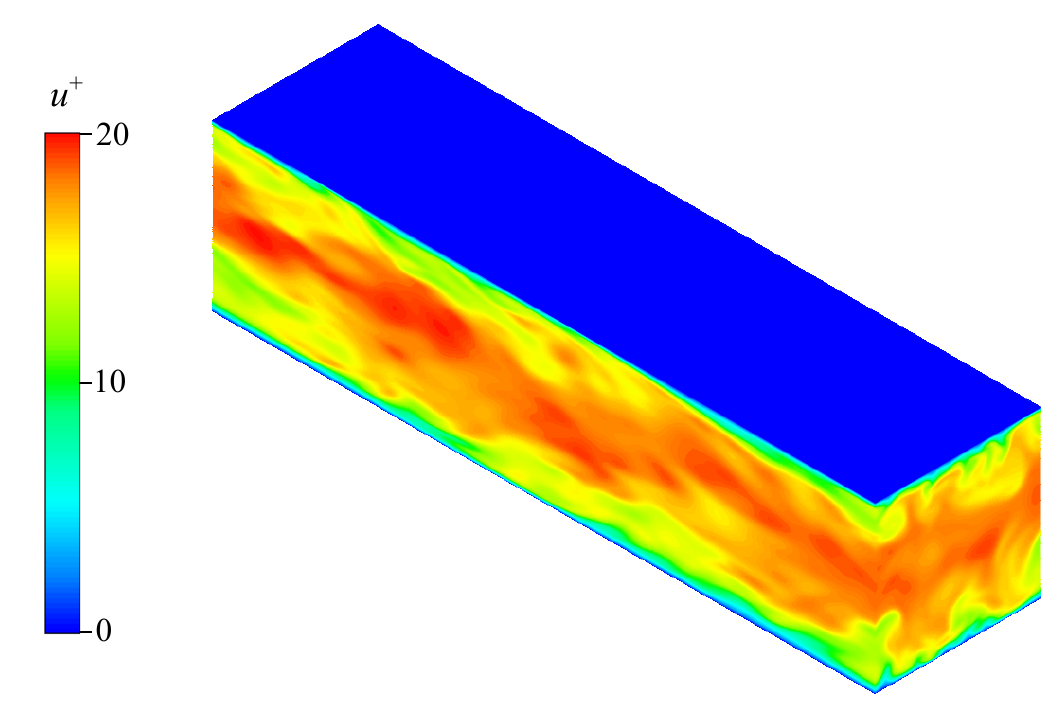}
\caption{Instantaneous streamwise velocity field obtained from the present gas-layer DNS for the stationary-interface case, $Re_{\tau}=180$ and $u_I^+=0$.}
\label{fig:dns_cnt}
\end{figure}

The DNS calculations are carried out using ANSYS Fluent by solving the incompressible Navier--Stokes equations without a turbulence model. The mesh is designed to resolve the relevant near-wall turbulent scales. In the present simulations, the grid spacings are $\Delta x^+=10$ in the streamwise direction and $\Delta z^+=7$ in the spanwise direction, while the wall-normal spacing is non-uniform and ranges from $\Delta y^+=0.1$ near the boundaries to approximately $\Delta y^+=4$ away from them. The final mesh contains $3.79\times10^6$ cells. The simulations are advanced until a statistically stationary turbulent state is reached, after which the mean velocity, root-mean-square velocity fluctuations and Reynolds shear stress are obtained by temporal averaging and by averaging over the homogeneous $x$--$z$ planes. An instantaneous streamwise velocity field from the statistically stationary state is shown in figure~\ref{fig:dns_cnt} for the stationary-interface case. The velocity is larger in the core of the gas layer and decreases towards both the prescribed-velocity lower boundary and the upper wall, as expected for a pressure-driven channel-like flow.

\begin{figure*}[!t]
\centering
\includegraphics[width=1.0\linewidth]{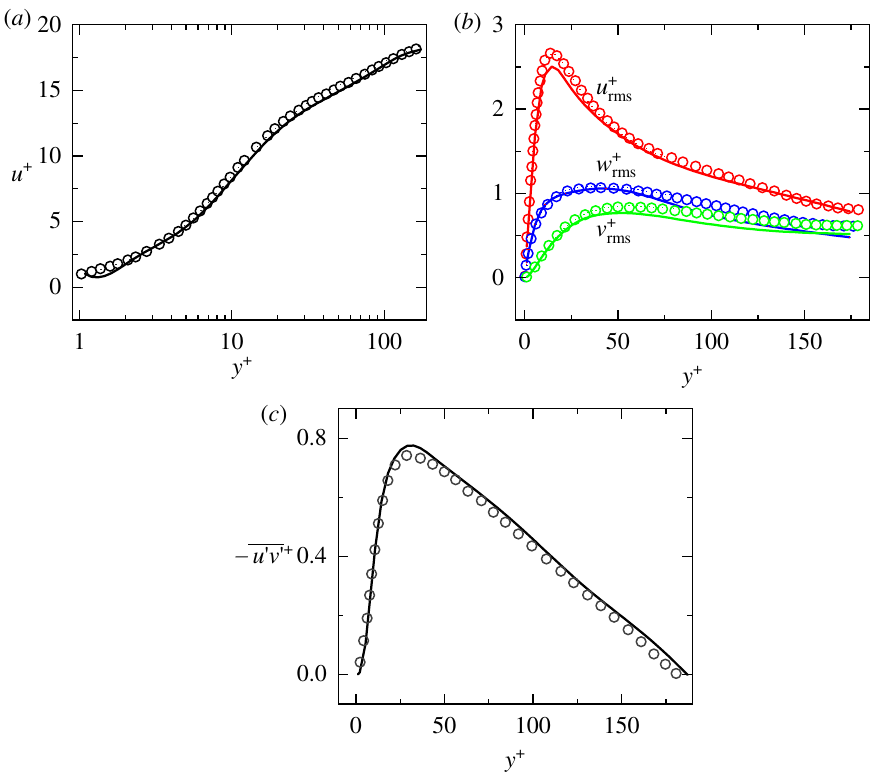}
\caption{Validation of the present gas-layer DNS against the reference DNS data of \citet{kim1987turbulence} for the stationary-interface case, $Re_{\tau}=180$ and $u_I^+=0$. In all panels, solid lines denote the present DNS results and open circles denote the reference data of \citet{kim1987turbulence}. ($a$) Mean streamwise velocity $u^+$ as a function of the wall coordinate $y^+$ measured from the stationary lower boundary. ($b$) Root-mean-square velocity fluctuations $u_{\mathrm{rms}}^+$, $v_{\mathrm{rms}}^+$ and $w_{\mathrm{rms}}^+$. ($c$) Reynolds shear stress $-\overline{u'v'}^+=-\overline{u'v'}/u_{\tau,p}^2$.}
\label{fig:dns_benchmark}
\end{figure*}

Before using the DNS data for the assessment of the interfacial damping models, we first validate the present DNS statistics against the reference channel-flow data of \citet{kim1987turbulence}. This comparison is carried out for the stationary-interface case, $u_I^+=0$, for which the gas-layer configuration reduces to the standard turbulent channel flow at $Re_{\tau}=180$. Figure~\ref{fig:dns_benchmark} shows the comparison for the mean streamwise velocity, the root-mean-square velocity fluctuations and the Reynolds shear stress. Here, $u_{\mathrm{rms}}^+=(\overline{u'u'})^{1/2}/u_{\tau,p}$, with analogous definitions for $v_{\mathrm{rms}}^+$ and $w_{\mathrm{rms}}^+$. The agreement is very good for all three statistics. In particular, the correct ordering of the fluctuation components, $u_{\mathrm{rms}}^+>w_{\mathrm{rms}}^+>v_{\mathrm{rms}}^+$, is reproduced, and the profile of the Reynolds shear stress component $-\overline{u'v'}^+=-\overline{u'v'}/u_{\tau,p}^2$ is also captured accurately. This comparison confirms the validity of the present DNS methodology for the channel-flow configuration considered here.

For the moving-interface cases, the same numerical procedure and mesh design are used, while the lower boundary velocity is set to the prescribed interfacial velocity. The resulting gas-layer DNS data are then used as reference solutions for the RANS/VOF calculations discussed below.

\subsection{Two-phase RANS/VOF implementation and validation cases}
\label{sec:rans_vof}

In contrast to the reference DNS, where only the gas layer is resolved and the interface is represented by a prescribed-velocity boundary, the RANS simulations include both the gas and liquid layers. The two-phase flow is computed in a one-field Eulerian framework using the coupled level-set and volume-of-fluid (CLSVOF) method available in ANSYS Fluent. The gas volume fraction is denoted by $\alpha_g$, and the liquid volume fraction is $\alpha_l=1-\alpha_g$. In the present implementation, the level-set distance field provides the geometrical interface representation used to evaluate interface normals and the wall-law distance $d_P$. The $\alpha_g=0.5$ iso-surface extracted from the VOF field is used for post-processing diagnostics, such as the evaluation of the interfacial velocity and the a posteriori flatness check described below. In this one-field formulation, a single velocity field and a single pressure field are solved over the whole domain, while the material properties are evaluated from the local volume fraction. Specifically,
\begin{equation}
\label{eq:mixture_properties}
\rho=\alpha_g\rho_g+(1-\alpha_g)\rho_l,\qquad
\mu=\alpha_g\mu_g+(1-\alpha_g)\mu_l .
\end{equation}

The flow geometry, domain size and boundary conditions are those described in \S~\ref{sec:canonical_configuration}: periodicity is imposed in the streamwise and spanwise directions, the upper gas-side boundary is a no-slip wall, and the lower liquid-side boundary is also treated as a no-slip wall. The same streamwise pressure gradient, $G=-\mathrm{d}p/\mathrm{d}x$, is imposed as in the DNS reference calculations. To reproduce the prescribed interfacial velocity used in the reference DNS, an additional streamwise body-force term is applied to the liquid phase. In practice, this forcing is weighted by the local liquid volume fraction, so that it acts only in cells where $\alpha_l$ is non-zero. Its magnitude is adjusted until the statistically steady interfacial velocity reaches the target value. The resulting interfacial velocity is evaluated from the mean streamwise velocity interpolated to the $\alpha_g=0.5$ iso-surface extracted from the VOF field. 

To suppress interface deformation and maintain the smooth-interface configuration, the surface tension is set to $\sigma=0.5~\mathrm{N\,m^{-1}}$. This value is chosen as a numerical constraint to keep the interface nearly flat, rather than as a parameter intended to reproduce a specific air--liquid-metal surface tension. With this value, capillary restoring effects are large compared with the air-side shear stress. Using the pressure-gradient friction velocity, the shear-based Weber number is $We_\tau=\rho_g u_{\tau,p}^2 H/\sigma\approx1.3\times10^{-2}$. This small value indicates that capillary forces strongly resist interface deformation at the gas-layer scale. The flatness of the interface was also checked a posteriori by monitoring the $\alpha_g=0.5$ iso-surface extracted from the VOF field. In all RANS cases considered here, the maximum interface displacement remained smaller than one local interface-normal cell thickness. The interface may therefore be regarded as flat for the purpose of the present smooth-interface benchmark.

The turbulence field is modelled using the low-Reynolds-number Wilcox-type $k$--$\omega$ closure in the same one-field framework. The transport equation for $k$ is solved without an additional interfacial source term,
\begin{equation}
\label{eq:k_transport_rans}
\rho\frac{\mathrm{D}k}{\mathrm{D}t}
=
P_k-\beta^\ast\rho k\omega
+
\frac{\partial}{\partial x_j}
\left[
\left(\mu+\sigma_k\mu_t\right)
\frac{\partial k}{\partial x_j}
\right].
\end{equation}
The corresponding $\omega$ equation is given in Eq.~\eqref{eq:omega_transport_model}, where the interfacial source term $S_\omega^I$ is determined either by the Egorov model or by the proposed target-$\omega$ formulation. \textcolor{black}{For the present formulation, the gas-side source localisation follows the pure-gas indicator $\chi_g$ defined in \S~\ref{sec:model}.} Although the turbulence equations are written in one-field form, the target value is constructed from gas-side quantities, while the interfacial source is localised on the gas side of the reconstructed interface, since the present study focuses on damping turbulence in the less dense phase. 

The production term is evaluated as $P_k=2\mu_tS_{ij}S_{ij}$, where $S_{ij}=(\partial u_i/\partial x_j+\partial u_j/\partial x_i)/2$ is the mean strain-rate tensor. The turbulent dynamic viscosity is computed as
\begin{equation}
\label{eq:turbulent_viscosity}
\mu_t
=
f_\mu\rho\frac{k}{\omega}.
\end{equation}
Here, $f_\mu$ is the low-Reynolds-number correction coefficient in the standard Fluent implementation of the Wilcox $k$--$\omega$ model. Following the Fluent default formulation, $f_\mu=(\alpha_0^\ast+Re_t/R_k)/(1+Re_t/R_k)$, where $Re_t=\rho k/(\mu\omega)$, $R_k=6$ and $\alpha_0^\ast=\beta/3$. Thus, $f_\mu\rightarrow1$ for large $Re_t$. The model constants used in the low-Reynolds-number Wilcox $k$--$\omega$ closure are $\beta^\ast=0.09$, $\beta=0.072$, $\alpha=0.52$, $\sigma_k=2.0$ and $\sigma_\omega=2.0$.

\textcolor{black}{Two additional numerical treatments are employed in the implementation to improve numerical robustness. For the discretisation of the diffusive flux in the $\omega$ equation, the molecular dynamic viscosity $\mu$ at faces crossing the gas--liquid interface is evaluated by harmonic rather than arithmetic averaging. This treatment limits the overestimation of the molecular diffusion coefficient associated with the large viscosity contrast between the two phases. In addition, a supplementary $\omega$ source based on the viscous-limit scaling is applied in the low-Reynolds-number liquid region to accelerate convergence. This treatment serves as a convergence aid and is kept unchanged in all calculations.}

\comm{
\textcolor{black}{In addition to the gas-side interfacial correction, a separate liquid-side damping treatment is used in the one-field RANS/VOF calculations to maintain the low turbulence level of the high-viscosity liquid. In the present implementation, this contribution is applied outside the numerically pure-gas region according to}
\begin{equation}
\label{eq:liquid_omega_source}
{\color{black}
S_{\omega,l}
=
(1-\chi_g)\beta\rho\omega_l^2,
}
\end{equation}
\textcolor{black}{where}
\begin{equation}
\label{eq:liquid_omega_target}
{\color{black}
\omega_l
=
\frac{6\mu_g}{\beta\rho_g d_P^2}.
}
\end{equation}
\textcolor{black}{This liquid-side term is used as a numerical treatment of the predominantly laminar liquid region and is kept unchanged in all calculations; the target-$\omega$ model assessed below refers specifically to the gas-side interfacial correction defined in \S~\ref{sec:model}.}
}
Two interfacial turbulence treatments are considered. The first is the Egorov-type damping model, in which a wall-like source term is added to the one-field $\omega$ equation. The second is the proposed target-$\omega$ formulation described in \S~\ref{sec:model}. In both cases, the source term is activated only in the gas-side interfacial region. For the Egorov model, the damping factor is varied over
\begin{equation}
\label{eq:egorov_B_values}
B=0.01,\ 0.1,\ 1,\ 10,\ 100 .
\end{equation}
\textcolor{black}{The target-$\omega$ formulation uses the fixed constants given in \S~\ref{sec:model}, with no case-specific parameter adjustment.}

\textcolor{black}{To assess the sensitivity to near-interface grid resolution, four meshes are considered. They are labelled by the interface-normal mesh spacing expressed in the pressure-gradient wall units defined in \S~\ref{sec:canonical_configuration},}
\begin{equation}
\label{eq:y_grid_plus}
{\color{black}
y_{\mathrm{grid}}^+
=
\frac{\rho_g u_{\tau,p}\Delta_\perp}{\mu_g},
}
\end{equation}
\textcolor{black}{where $\Delta_\perp$ is the local mesh thickness normal to the interface. The values considered are}
\begin{equation}
\label{eq:y_grid_values}
y_{\mathrm{grid}}^+=0.5,\ 1,\ 3,\ 6 .
\end{equation}
\textcolor{black}{For the flat, mesh-aligned canonical configuration, the centre of the first pure-gas cell is located at $d_P=\Delta_\perp/2$. The nominal grid measure $y_{\mathrm{grid}}^+$ should therefore be distinguished from the model-internal coordinate $d_P^\ast=\rho_g u_\tau^{\mathrm{eff}}d_P/\mu_g$ introduced in \S~\ref{sec:model}: the former characterises the prescribed mesh resolution using $u_{\tau,p}$, whereas the latter varies with the locally determined effective friction velocity.}

\textcolor{black}{The same target-$\omega$ formulation is evaluated at all three prescribed interfacial velocities, $u_I^+=0$, $0.95$ and $1.9$, so that each DNS case provides a validation of the model at a different interfacial velocity.} The resulting gas-side mean velocity, turbulent kinetic energy and Reynolds shear-stress profiles are compared with the DNS reference data in \S~\ref{sec:results}, with particular attention to the sensitivity to $y_{\mathrm{grid}}^+$ and to the Egorov damping factor $B$.

\textcolor{black}{An overall solver-level timing comparison was performed for the canonical case with $u_I^+=1.9$ and $y_{\mathrm{grid}}^+=3$. For both damping formulations, the calculations were started from the same initial condition and advanced to the final steady converged state. Five repeated calculations were performed using the present formulation and another five using the Egorov model with $B=1$, with identical solver settings. The convergence criteria were set to $10^{-6}$ for the continuity equation and $10^{-3}$ for all other governing equations. Using 32 Fluent solver processes on an AMD Ryzen~9~7950X processor, the mean elapsed times were $512.53~\mathrm{s}$ and $524.38~\mathrm{s}$, respectively. Thus, although the present formulation involves an additional local Newton inversion in the activated interfacial cells, no appreciable increase in the overall wall-clock elapsed time was observed in this complete steady benchmark; the two mean elapsed times differed by only about $2.3\%$. This solver-level timing comparison includes both local source-term evaluation and the overall convergence history, and should not be interpreted as an isolated measurement of the intrinsic cost of the wall-law inversion.}

\section{Results and discussion}
\label{sec:results}

The target-$\omega$ formulation, hereafter referred to as the present model, is assessed first in the canonical smooth-interface configuration introduced in \S~\ref{sec:configuration}. For both the RANS and DNS simulations, the friction Reynolds number is fixed at $Re_\tau\simeq180$, while \textcolor{black}{three} interfacial velocities are considered: $u_I^+=0$, \textcolor{black}{$0.95$} and $1.9$. Further assessments are subsequently performed against the stratified-flow experiments of \citet{fabre1987experimental} and \citet{staebler2007}.

\subsection{\textcolor{black}{DNS-based validation in the canonical smooth-interface configuration}}
\label{sec:results_dns_validation}

We first consider the stationary-interface case, $u_I^+=0$, and the faster moving-interface case, $u_I^+=1.9$. For these two cases, the present predictions are compared with those obtained using the Egorov-type model over different near-interface grid resolutions, while the sensitivity of the Egorov model to the damping factor $B$ is also examined. \textcolor{black}{The intermediate case, $u_I^+=0.95$, is then considered as an additional DNS-based validation of the present formulation.} For all three cases, the streamwise and spanwise directions are homogeneous. The reported quantities are therefore statistically averaged over planes of constant $y$. The profiles are plotted against $y/H$, which increases from 0 at the gas--liquid interface to 1 at the upper wall. \textcolor{black}{The DNS reference values used in the comparisons are shown as discrete open-circle samples at the wall-normal locations available for the present comparison.}

\begin{figure*}[!t]
\centering
\includegraphics[width=1.0\linewidth]{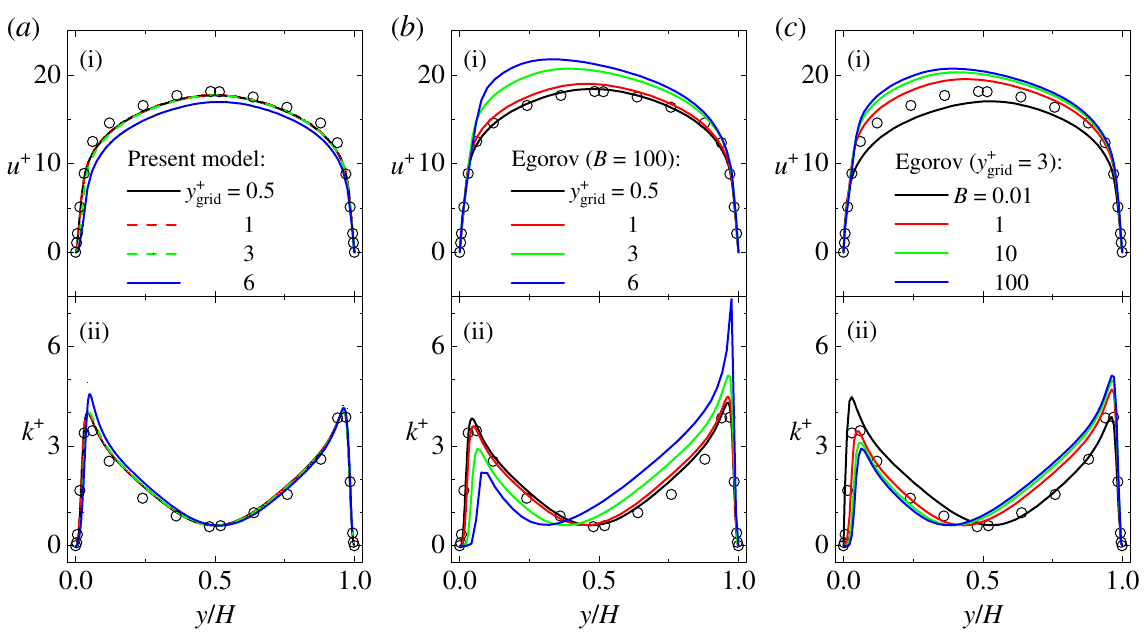}
\caption{Comparison between RANS predictions and DNS results for the stationary-interface case, $u_I^+=0$. Open circles denote DNS reference samples, and lines denote RANS predictions. ($a$) Present model for different values of $y_{\mathrm{grid}}^+$. ($b$) Egorov model with $B=100$ for different values of $y_{\mathrm{grid}}^+$. ($c$) Egorov model at fixed $y_{\mathrm{grid}}^+=3$ for different values of $B$. Row ($i$): mean streamwise velocity $u^+$. Row ($ii$): turbulent kinetic energy $k^+$.}
\label{fig:u_k_ui0}
\end{figure*}

Figures~\ref{fig:u_k_ui0}($a$-$i$) and \ref{fig:u_k_ui0}($a$-$ii$) show, respectively, the mean streamwise velocity $u^+$ and the turbulent kinetic energy $k^+$ predicted by the present model for the stationary-interface case, $u_I^+=0$. Results obtained on grids of different near-interface resolutions, characterised by $y_{\mathrm{grid}}^+$, are compared with the corresponding DNS data. The predictions obtained with $y_{\mathrm{grid}}^+=0.5$, $1$ and $3$ remain close to one another for both quantities. A more noticeable grid dependence appears on the coarsest grid, $y_{\mathrm{grid}}^+=6$, particularly in the near-boundary peaks of $k^+$ and, to a lesser extent, in the mean velocity. Nevertheless, the overall velocity distribution and the characteristic two-peak structure of $k^+$ remain well reproduced. In the bulk region, roughly $0.1\lesssim y/H\lesssim0.9$, the deviation in $u^+$ remains within approximately \textcolor{black}{$15\%$} for all grid resolutions. The largest discrepancy in $k^+$ occurs near the boundaries on the coarsest grid, where the peak deviation is approximately \textcolor{black}{$23\%$}.

For comparison, figures~\ref{fig:u_k_ui0}($b$)--\ref{fig:u_k_ui0}($c$) show the corresponding predictions obtained with the Egorov model under different parametric conditions. Let us first consider the series of simulations performed at fixed grid resolution, $y_{\mathrm{grid}}^+=3$, while varying $B$; see figures~\ref{fig:u_k_ui0}($c$-$i$) and \ref{fig:u_k_ui0}($c$-$ii$). Since the imposed specific dissipation rate in the Egorov model is proportional to $B$, too small a value of $B$ weakens the interfacial turbulence damping and may therefore lead to an overestimation of the turbulent kinetic energy near the interface. This is approximately what is observed for $B=0.01$: compared with the DNS data, the peak of $k^+$ on the interface-facing side is overestimated by about $22\%$. At the same time, the mean streamwise velocity is clearly underestimated on the interface-facing side, with a typical deviation of about $15\%$. This underestimation is consistent with an excessive turbulent viscosity on the interface-facing side. In the eddy-viscosity framework, such an overprediction tends to enhance turbulent momentum transport and to smooth the mean velocity gradient, thereby reducing the predicted value of $u^+$. This interpretation is also supported by the modelled turbulent shear stress discussed below in figure~\ref{fig:shear_rs_ui02}, where the small-$B$ Egorov case yields an overly large turbulent momentum transport near the interface. As $B$ is increased, the predicted $k^+$ on the interface-facing side decreases, whereas the opposite trend is observed for $u^+$. Among the values of $B$ tested here, the best overall agreement on this grid is obtained for $B\simeq1$, which is two orders of magnitude smaller than the value $B=100$ used by \citet{egorov2004} in their original free-surface source-term implementation.

A different behaviour is observed when $B$ is fixed at $100$ while the near-interface grid resolution is varied. As figures~\ref{fig:u_k_ui0}($b$-$i$) and \ref{fig:u_k_ui0}($b$-$ii$) show, the predictions become progressively more sensitive to the grid spacing as $y_{\mathrm{grid}}^+$ is increased. On the coarsest grid, the maximum deviation in $u^+$ reaches approximately $30\%$, while the wall-side peak of $k^+$ is overestimated by approximately $85\%$. The response reflects both the explicit dependence of the Egorov target value on the interface-normal mesh spacing and changes in the resolved near-interface velocity gradient and turbulence production as the mesh is coarsened. The comparison therefore shows that the effective response of the Egorov formulation depends jointly on the prescribed damping factor $B$ and the near-interface grid resolution.

\begin{figure*}[!t]
\centering
\includegraphics[width=1.0\linewidth]{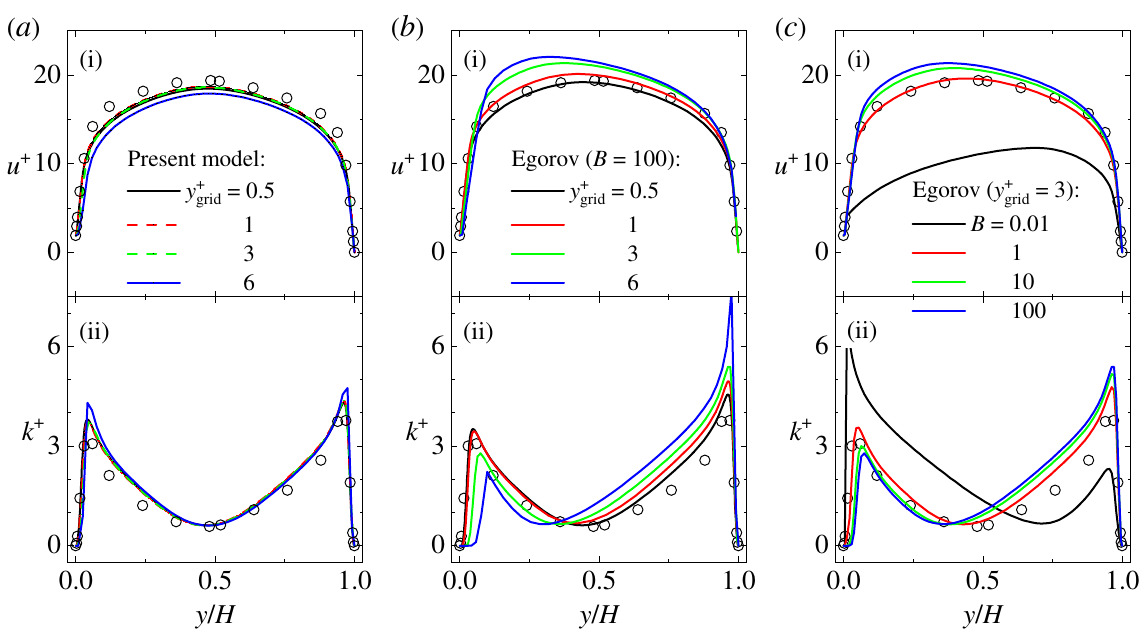}
\caption{Same as figure~\ref{fig:u_k_ui0}, but for the moving-interface case, $u_I^+=1.9$.}
\label{fig:u_k_ui19}
\end{figure*}

Figure~\ref{fig:u_k_ui19} provides the same comparison as figure~\ref{fig:u_k_ui0}, but for the moving-interface case, $u_I^+=1.9$. Owing to the finite interface velocity, the profiles of $u^+$ and $k^+$ are no longer symmetric with respect to $y/H=0.5$. According to the DNS results, the wall-normal position of the peak in $u^+$ is slightly shifted towards the interface-facing side. For $k^+$, the interface-facing and wall-facing peaks also become asymmetric.

These changes are reproduced by the present model; see figures~\ref{fig:u_k_ui19}($a$-$i$) and \ref{fig:u_k_ui19}($a$-$ii$). As in the stationary-interface case, the predictions on the three finer meshes remain close to one another, while the coarsest grid produces the largest departure, particularly in the near-boundary values of $k^+$. The asymmetric mean-velocity and turbulence distributions associated with the moving interface are nevertheless retained over all four grid resolutions. The deviation in $u^+$ remains within approximately \textcolor{black}{$20\%$}, while the largest near-boundary deviation in $k^+$ on the coarsest grid is approximately \textcolor{black}{$37\%$}.

For the Egorov model, the sensitivity to both the prefactor $B$ and the grid resolution $y_{\mathrm{grid}}^+$ remains evident. At fixed $y_{\mathrm{grid}}^+=3$, figures~\ref{fig:u_k_ui19}($c$-$i$) and \ref{fig:u_k_ui19}($c$-$ii$) show that the prediction deteriorates substantially when $B$ is reduced to the very small value $B=0.01$. At the opposite limit, $B=100$, the imposed damping produces an interface-side peak of $k^+$ that differs from DNS by approximately $13\%$, while the near-wall peak is overestimated by approximately $42\%$. The simulations performed at fixed $B=100$ also exhibit pronounced grid dependence. On the coarsest grid, $y_{\mathrm{grid}}^+=6$, the interface--wall asymmetry of the $k^+$ profile is amplified: the interface-side peak differs from DNS by approximately $30\%$, whereas the wall-side peak is overestimated by approximately $90\%$. These trends are consistent with those observed for the stationary-interface case and demonstrate that the Egorov response remains strongly coupled to both $B$ and the prescribed mesh resolution.

As a supplementary comparison, figure~\ref{fig:shear_rs_ui02} shows the Reynolds shear-stress component obtained from the DNS and the corresponding turbulent shear stress predicted by the RANS eddy-viscosity closure. The DNS quantity is the normalised kinematic Reynolds shear stress $-\overline{u'v'}^+=-\overline{u'v'}/u_{\tau,p}^2$. In the RANS calculations, this quantity is not directly resolved, but is modelled through the Boussinesq eddy-viscosity approximation. For the present plane-averaged profiles, the corresponding modelled contribution is
\begin{equation}
\label{eq:rs_rans}
-\overline{u'v'}_{\mathrm{RANS}}^+
=
\frac{2\mu_tS_{xy}}{\rho_g u_{\tau,p}^2},
\end{equation}
where $S_{xy}$ is the mean strain-rate component and the quantities are evaluated in the gas layer. This comparison provides an additional assessment of the turbulent momentum transport implied by the eddy-viscosity field and the mean velocity gradient.

\begin{figure*}[!t]
\centering
\includegraphics[width=1.0\linewidth]{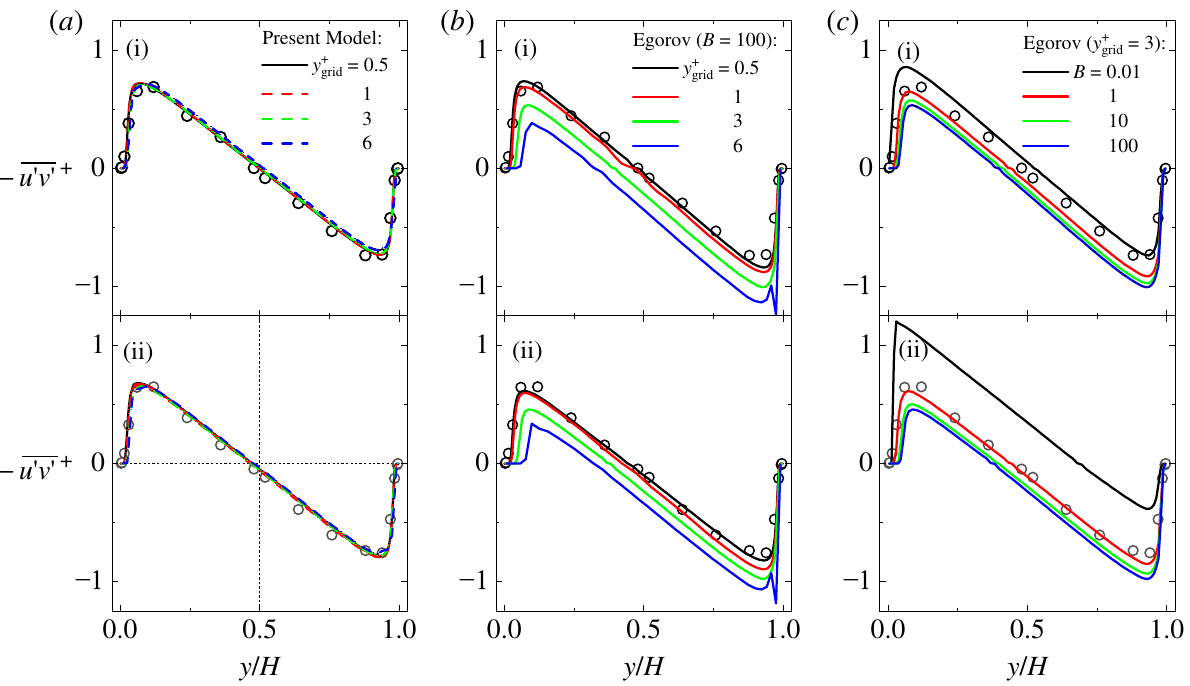}
\caption{Comparison between the Reynolds shear-stress component obtained from DNS and the corresponding turbulent shear stress predicted by the RANS eddy-viscosity closure for \textcolor{black}{the two interface-velocity cases}. Row ($i$): $u_I^+=0$. Row ($ii$): $u_I^+=1.9$. Open circles denote the DNS quantity $-\overline{u'v'}^+$, while lines denote the RANS predictions obtained from the Boussinesq eddy-viscosity approximation. ($a$): Present model for different values of $y_{\mathrm{grid}}^+$. ($b$): Egorov model with $B=100$ for different values of $y_{\mathrm{grid}}^+$. ($c$): Egorov model at fixed $y_{\mathrm{grid}}^+=3$ for different values of $B$. The dashed lines in panel ($a$-$ii$) mark $y/H=0.5$ and $-\overline{u'v'}^+=0$, highlighting the loss of mid-height antisymmetry in the moving-interface case.}
\label{fig:shear_rs_ui02}
\end{figure*}

The DNS profiles in figure~\ref{fig:shear_rs_ui02} reveal how the Reynolds shear-stress distribution changes when interfacial motion is imposed. For the stationary-interface case, $u_I^+=0$, the DNS profile is antisymmetric with respect to the channel mid-height. For the moving-interface case, $u_I^+=1.9$, this mid-height antisymmetry is lost: the Reynolds shear-stress component no longer vanishes at $y/H=0.5$, as highlighted by the dashed lines in figure~\ref{fig:shear_rs_ui02}($a$-$ii$). The present model captures both the antisymmetric distribution in the stationary-interface case and its loss in the moving-interface case. The four present-model profiles remain close to one another over most of the gas layer, whereas the Egorov predictions exhibit a substantially stronger dependence on both $y_{\mathrm{grid}}^+$ and $B$, consistent with the behaviour observed above for $u^+$ and $k^+$.

Before closing \textcolor{black}{the direct comparison between the two damping formulations}, we note that the sensitivity of the Egorov model to grid resolution and to the prefactor $B$ is not specific to the smooth channel flows considered here. Similar grid- and parameter-dependent behaviour has been reported in previous studies of stratified channel flows involving small interfacial waves; see, e.g., \citet{frederix2018}, \citet{fan2019} and \citet{polansky2022implementation}.

\textcolor{black}{We next consider the intermediate interface velocity, $u_I^+=0.95$, as an additional DNS-based validation of the present formulation. No Egorov calculations are included for this case; the purpose is to examine whether the present model retains consistent behaviour between the two interface-velocity conditions considered above.}

\begin{figure*}[!t]
\centering
\includegraphics[width=1.0\linewidth]{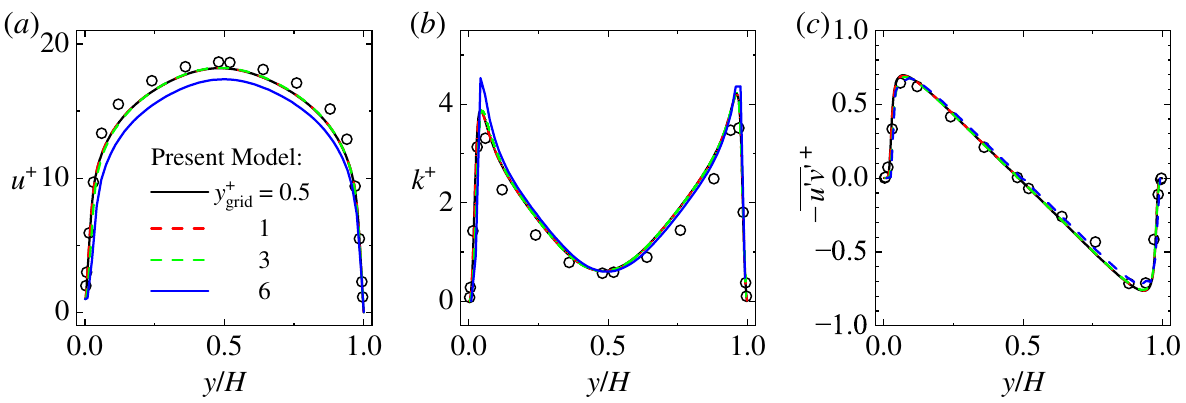}
\caption{\textcolor{black}{Validation} of the present model for the intermediate moving-interface case, $u_I^+=0.95$. Open circles denote DNS reference samples, and lines denote RANS predictions obtained with the present model for different values of $y_{\mathrm{grid}}^+$. ($a$) Mean streamwise velocity $u^+$. ($b$) Turbulent kinetic energy $k^+$. ($c$) Reynolds shear-stress component $-\overline{u'v'}^+=-\overline{u'v'}/u_{\tau,p}^2$.}
\label{fig:u_k_rs_ui1}
\end{figure*}

Figure~\ref{fig:u_k_rs_ui1} compares the DNS results with the present-model predictions for the intermediate interface velocity. The mean streamwise velocity, shown in figure~\ref{fig:u_k_rs_ui1}($a$), is reproduced well over the gas layer. The predictions obtained with $y_{\mathrm{grid}}^+=0.5$, $1$ and $3$ remain close, whereas the coarsest grid gives a somewhat lower velocity over much of the layer. The overall profile shape nevertheless remains consistent with the DNS, with the deviation in the bulk region remaining within approximately \textcolor{black}{$17\%$}.

The turbulent kinetic energy profiles are shown in figure~\ref{fig:u_k_rs_ui1}($b$). The DNS reference samples exhibit two near-boundary peaks separated by a minimum in the central part of the gas layer. The present model reproduces this overall structure for all four grid resolutions. Grid dependence is again most pronounced on the coarsest mesh, where the near-boundary peaks are more strongly overpredicted. \textcolor{black}{On the coarsest mesh, the near-interface RANS peak is approximately $37\%$ higher than the largest available DNS reference sample in that region. Since the DNS reference data used for this comparison are available at discrete wall-normal sampling locations, this value should be regarded as an approximate peak-level discrepancy rather than an exact error in the continuous peak magnitude.} In contrast, the variation of $k^+$ in the central part of the gas layer remains comparatively insensitive to the near-interface grid resolution.

Figure~\ref{fig:u_k_rs_ui1}($c$) shows the corresponding Reynolds shear-stress component. The four RANS profiles nearly collapse and follow the DNS distribution closely over most of the gas layer, including the change of sign between the interface-facing and wall-facing regions. The discrepancy at the characteristic extrema remains within approximately \textcolor{black}{$5\%$}. The good agreement in the shear stress, despite the larger local deviations in the near-boundary $k^+$ peaks on the coarsest mesh, indicates that the predicted turbulent momentum transport is less sensitive to the near-interface grid resolution than the local turbulence-energy peaks.

\textcolor{black}{Taken together, the three canonical cases show that the present target-$\omega$ formulation provides consistent predictions over $0\le u_I^+\le1.9$ without adjustment of its fixed constants. The mean velocity and Reynolds shear-stress distributions are comparatively insensitive to the near-interface grid resolution, while the main discrepancies occur in the near-boundary turbulence levels on the coarsest mesh. A non-negligible residual grid dependence therefore remains, especially for the near-boundary turbulence-energy peaks on the coarsest mesh. Unlike the Egorov prescription, the present source strength does not contain the same explicit inverse-square dependence on the interface-normal mesh spacing. However, the geometric distance $d_P$ enters the effective wall coordinate $d_P^\ast$ and hence the blended target value $\omega_{\mathrm{tar}}$, so changes in the near-interface mesh modify the wall-law sampling location and can affect the predicted near-boundary turbulence level, as seen most clearly for $y_{\mathrm{grid}}^+=6$.}

\subsection{Assessment against the Fabre et al. stratified-flow experiment}
\label{sec:results_fabre}

We next assess the proposed model against the stratified air--water experiment of \citet{fabre1987experimental}, focusing on case ``250''. This experiment has been widely used as a benchmark for interfacial turbulence-damping models in Reynolds-averaged simulations of stratified two-phase flow; see, e.g., \citet{yao2005three}, \citet{hohne2014validation}, \citet{frederix2018} and \citet{dong2022}.

\begin{figure}[!tbp]
\centering
\includegraphics[width=\linewidth]{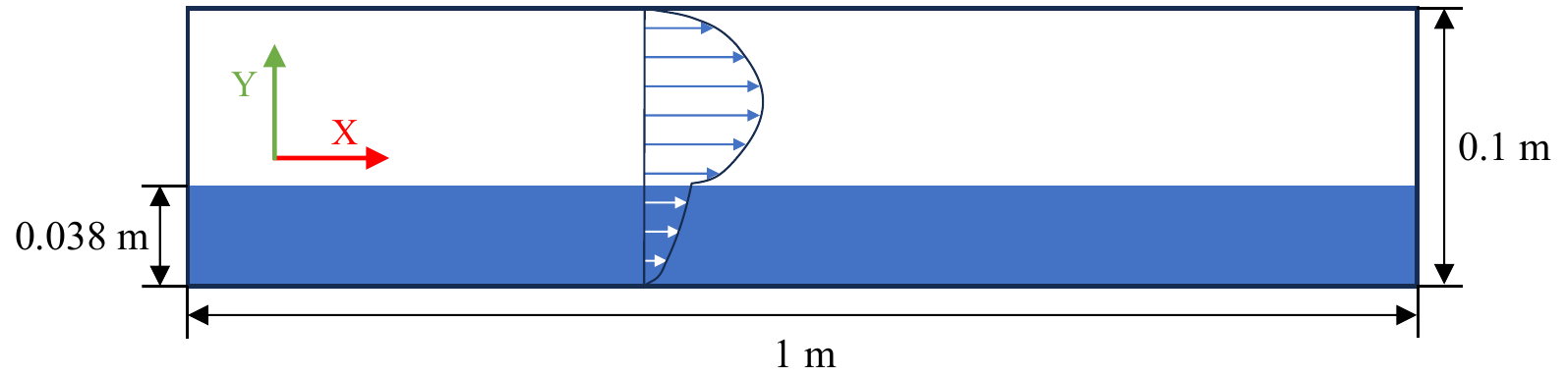}
\caption{Schematic of the two-dimensional computational domain used for the \citet{fabre1987experimental} stratified-flow case.}
\label{fig:problem_fabre}
\end{figure}

The computational setup follows the periodic pressure-gradient-driven configuration used in previous numerical studies of this experiment \citep{frederix2018,dong2022}. A two-dimensional domain of length $1~\mathrm{m}$ and height $0.1~\mathrm{m}$ is considered, as sketched in figure~\ref{fig:problem_fabre}. The liquid layer occupies the lower part of the channel, with a mean depth of $0.038~\mathrm{m}$, leaving a gas-layer height of approximately $H_g=0.062~\mathrm{m}$. The upper and lower boundaries are treated as no-slip walls, and periodic boundary conditions are imposed in the streamwise direction. The flow is driven by an imposed pressure gradient, adjusted to reproduce the experimental flow rates of Fabre case ``250''. The working fluids are air and water. In the present Fluent calculation, the default standard air--water properties are used: $\rho_g=1.23~\mathrm{kg\,m^{-3}}$, $\mu_g=1.79\times10^{-5}~\mathrm{Pa\,s}$, $\rho_l=998~\mathrm{kg\,m^{-3}}$ and $\mu_l=1.00\times10^{-3}~\mathrm{Pa\,s}$.

For reference, the experimental data of \citet{fabre1987experimental} report a pressure-gradient magnitude of approximately $G=2.10~\mathrm{Pa\,m^{-1}}$ for case ``250''. We use this value to define a pressure-gradient friction velocity for comparison with the canonical cases. Using the gas-layer height $H_g=0.062~\mathrm{m}$ and the same pressure-gradient friction-velocity definition as in \S~\ref{sec:canonical_configuration}, one obtains $u_{\tau,p}\simeq0.23~\mathrm{m\,s^{-1}}$. The corresponding friction Reynolds number, defined with the gas-layer half-height as in the canonical configuration, is $Re_\tau = {\rho_g u_{\tau,p}(H_g/2)}/{\mu_g} \simeq500$. The measured liquid velocity near the interface is about $0.5~\mathrm{m\,s^{-1}}$, giving $u_I^+\simeq2.1$. Thus, the dimensionless interface velocity is close to that of the moving-interface \textcolor{black}{canonical case}, $u_I^+=1.9$, whereas the Reynolds number is higher than in the canonical DNS cases. \textcolor{black}{The Fabre case therefore provides a useful external assessment of the transferability of the present target-$\omega$ formulation beyond the canonical channel configuration.} The near-interface grid resolution in the present RANS calculation is $y_{\mathrm{grid}}^+\simeq{2.56}$, \textcolor{black}{evaluated from the interface-normal grid spacing $\Delta_\perp$ and $u_{\tau,p}$.}

To place the present results in the context of existing interfacial damping treatments, we also include reference predictions from the literature. The reference curves shown below are digitised from the comparative results reported by \citet{dong2022}. In their terminology, the DB-damping curve corresponds to the density-based damping model proposed by Dong et al., whereas the PL-damping curve corresponds to their implementation of the phenomenological length-scale damping approach of \citet{frederix2018}. A brief summary of these reference damping models is provided in \ref{app:damping_models}. The PL-damping curve is therefore used here as a Frederix-type calibrated reference result rather than as a new calculation performed in the present work. \citet{dong2022} noted that the PL-damping parameters were calibrated for this co-current Fabre case, whereas their DB-damping model does not rely on a case-specific empirical damping coefficient.

In contrast to the preceding subsections, where the results were presented in pressure-gradient wall units, all quantities in the present comparison are shown in dimensional form. This facilitates a direct comparison with the experimental measurements and with the reference numerical results reported in the literature. The profiles are also plotted with the wall-normal coordinate on the vertical axis and the flow quantity on the horizontal axis, so that the rapid variation of the flow statistics across the gas--liquid interface can be more clearly identified.

\begin{figure*}[!t]
\centering
\includegraphics[width=1.0\linewidth]{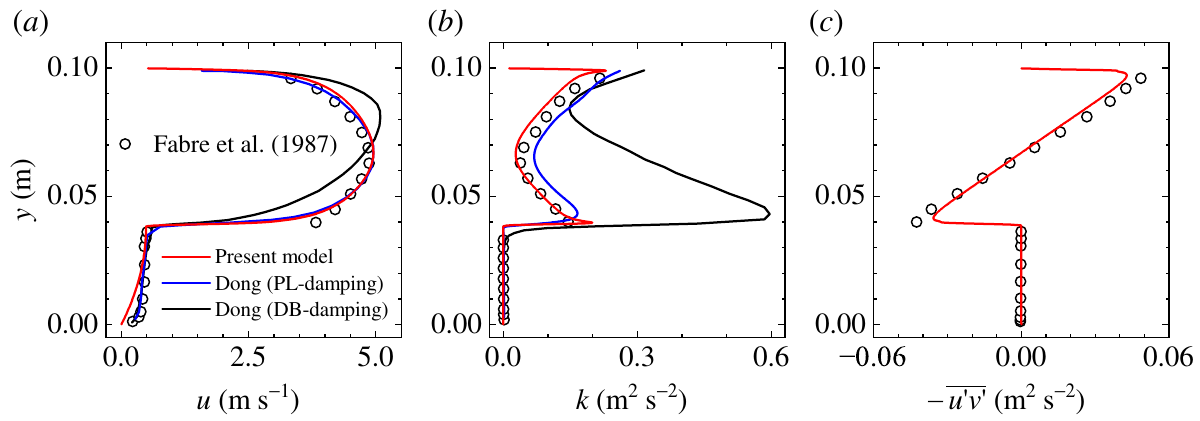}
\caption{Comparison between the experimental data of \citet{fabre1987experimental} and RANS predictions for stratified-flow case ``250''. Open circles denote the experimental data, and the red line denotes the present model. In panels ($a$) and ($b$), the blue and black curves are digitised from the comparative results of \citet{dong2022}: the blue line denotes their PL-damping result, corresponding to the phenomenological length-scale damping approach of \citet{frederix2018}, and the black line denotes their DB-damping result. ($a$) Mean streamwise velocity $u$. ($b$) Turbulent kinetic energy $k$. ($c$) Reynolds shear-stress component $-\overline{u'v'}$. Reference DB- and PL-damping curves are not shown in panel ($c$) because directly comparable Reynolds-stress data are not available from \citet{dong2022}.}
\label{fig:u_k_RS_fabre}
\end{figure*}

Figure~\ref{fig:u_k_RS_fabre} compares the experimental data of \citet{fabre1987experimental} with the present RANS prediction and with the two reference curves digitised from \citet{dong2022}. The mean streamwise velocity is shown in figure~\ref{fig:u_k_RS_fabre}($a$). The present model reproduces the experimental velocity profile well in the gas layer. In the gas-layer core, away from both the interface and the upper wall, the deviation from the experimental data remains within approximately $5\%$. The agreement is comparable to that obtained with the PL-damping curve, which is consistent with the fact that the PL-damping formulation represents a calibrated phenomenological damping treatment for this test case. By contrast, in the digitised comparison used here, the DB-damping curve shows a larger deviation in the gas-side velocity profile, with an underprediction near the interface and an overprediction towards the upper part of the gas layer.

The corresponding turbulent kinetic energy profiles are shown in figure~\ref{fig:u_k_RS_fabre}($b$). The experimental data exhibit a moderate level of gas-side turbulence, without the strong interfacial peak that would be obtained in the absence of sufficient turbulence damping. The present model captures this trend and gives a reasonable prediction throughout the gas layer. The turbulent kinetic energy is slightly overestimated near the interface, but its level in the gas-layer core remains close to the experimental data; the maximum discrepancy in the core region is about \textcolor{black}{$20$--$25\%$}. The PL-damping curve also gives good agreement for this case, again reflecting its case-specific calibration. In the digitised comparison used here, the DB-damping curve gives a substantially larger value of $k$ near the interface, suggesting weaker gas-side damping for this particular setup.

Figure~\ref{fig:u_k_RS_fabre}($c$) further compares the Reynolds shear-stress component $-\overline{u'v'}$. Unlike the comparisons of $u$ and $k$, only the present prediction is shown for $-\overline{u'v'}$, since the corresponding DB- and PL-damping results digitised from \citet{dong2022} are not available in a directly comparable form. The present model captures the main structure of the measured Reynolds shear-stress profile. In the liquid layer, $-\overline{u'v'}$ remains close to zero, consistent with the weak turbulence level measured below the interface. In the gas layer, both the sign change and the magnitude of $-\overline{u'v'}$ are reasonably reproduced. The agreement is particularly satisfactory in the gas-layer core, where the model follows the experimental trend from positive values near the interface to negative values closer to the upper wall. This comparison provides a more direct assessment of turbulent momentum transport than the $k$ profile alone, because the modelled Reynolds shear stress depends on the combined effect of the eddy viscosity and the mean velocity gradient.

Overall, the comparison with the Fabre et al. experiment shows that the proposed target-$\omega$ model provides a satisfactory prediction of the mean velocity and Reynolds shear stress, together with a reasonable prediction of the gas-side turbulent kinetic energy, in a large-scale stratified-flow configuration. \textcolor{black}{Importantly, no Fabre-specific adjustment of the fixed constants in the target-$\omega$ relation is introduced.} The remaining discrepancy is mainly associated with a moderate overprediction of $k$ near the interface, whereas the velocity and Reynolds shear-stress profiles in the gas layer are well reproduced.

\begin{revision}
\subsection{Assessment against the St\"abler counter-current stratified-flow experiment}
\label{sec:staebler}

As a further external assessment, the present model is applied to the counter-current air--water experiment of \citet{staebler2007} in the horizontal rectangular channel of the WENKA facility. In contrast to the canonical cases and the Fabre configuration considered above, the liquid level is not prescribed in the present calculation but develops as part of the coupled two-phase solution. Experimental case ``3'' is considered, corresponding to the measurement section MP1 located $235\,\mathrm{mm}$ downstream of the liquid inlet. This operating condition has also been used as a benchmark in previous numerical studies \citep{porombka2015,hohne2018,dong2022}.

The two-dimensional computational configuration is shown in figure~\ref{fig:staebler-config}. Following the central-section treatment adopted in previous numerical studies, both phases are included in the computational domain and the gas--liquid interface is captured using the CLSVOF method. The main WENKA test section has a height of $90\,\mathrm{mm}$, with a $9\,\mathrm{mm}$-high liquid inlet and an $81\,\mathrm{mm}$-high gas inlet. A fully developed gas-side velocity profile is prescribed at the gas inlet and normalised to give a two-dimensional gas unit discharge of $q_{\mathrm{in,G}}\simeq0.36\,\mathrm{m^2\,s^{-1}}$; the corresponding inlet turbulence quantities are prescribed together with this profile. On the liquid side, a uniform inlet velocity of $U_{\mathrm{in,L}}=0.7\,\mathrm{m\,s^{-1}}$ is imposed over the $9\,\mathrm{mm}$-high inlet, corresponding to $q_{\mathrm{in,L}}=0.0063\,\mathrm{m^2\,s^{-1}}$. An extended inlet section is used to allow the initially uniform liquid flow to develop before entering the main channel. A constant pressure $p_a$ is imposed at the outlet boundaries, and no-slip conditions are applied at the solid walls. The same fluid properties as in the Fabre assessment are used.

\begin{figure*}[!t]
    \centering
    \includegraphics[width=1.0\linewidth]{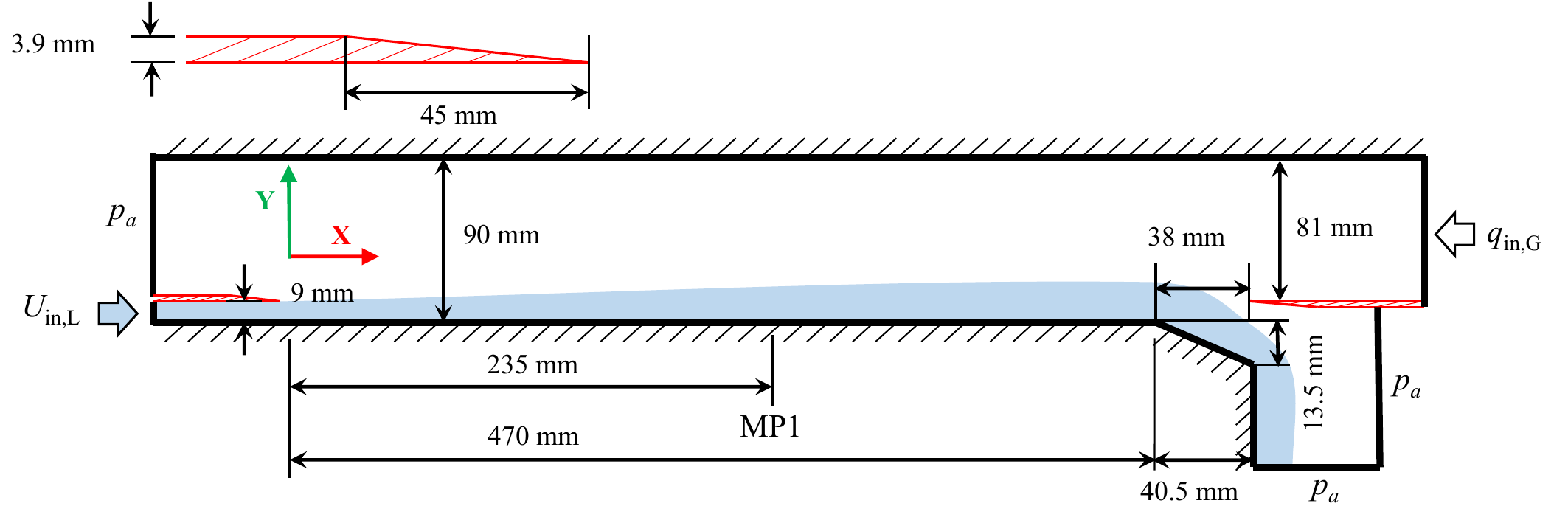}
    \caption{\textcolor{black}{Computational configuration used for the assessment against the counter-current stratified-flow experiment of St\"abler (2007). The MP1 section is located $235\,\mathrm{mm}$ downstream of the liquid inlet.}}
    \label{fig:staebler-config}
\end{figure*}

For case ``3'', the experimentally measured mean liquid depth at MP1 is $y_m=10.24\,\mathrm{mm}$ \citep{staebler2007}, corresponding to a mean gas-layer height of approximately $H_g=79.8\,\mathrm{mm}$. The hydraulic-diameter Reynolds number reported for the gas phase is $\simeq2.7\times10^4$ \citep{hohne2018}. The characteristic geometric and Reynolds-number scales therefore differ substantially from those of the canonical DNS cases. The computational mesh is uniform in the cross-stream direction, with a grid spacing of approximately $\Delta y=0.81\,\mathrm{mm}$. At MP1, this corresponds to $\Delta y/H_g\simeq1.0\times10^{-2}$ based on the experimentally measured mean gas-layer height. Since the present configuration is driven by counter-current inlet streams rather than by a prescribed uniform pressure gradient, a pressure-gradient friction velocity and the associated $y_{\mathrm{grid}}^+$ used for the canonical and Fabre cases are not defined here.

\begin{figure*}[!t]
    \centering
    \includegraphics[width=0.9\linewidth]{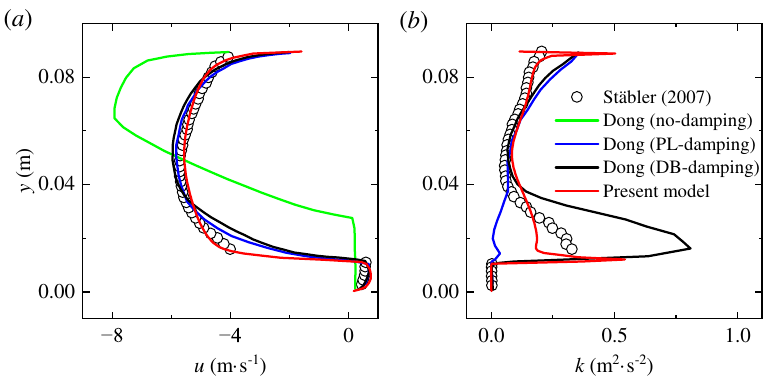}
    \caption{\textcolor{black}{Comparison between the experimental data of St\"abler (2007) and RANS predictions obtained using different interfacial turbulence-damping treatments for stratified-flow case ``3'' at MP1. Open circles denote the experimental measurements, and the red curve denotes the present target-$\omega$ model. The blue and black curves denote the PL-damping and DB-damping reference results, respectively. \textcolor{black}{In panel ($a$), the green curve denotes the no-damping reference result reported by \citet{dong2022}.} ($a$) Mean streamwise velocity. ($b$) Turbulent kinetic energy.}}
    \label{fig:staebler-results}
\end{figure*}

Figure~\ref{fig:staebler-results} compares the present predictions of mean streamwise velocity and turbulent kinetic energy at MP1 with the measurements of \citet{staebler2007}. Reference RANS predictions obtained using no interfacial damping, PL-damping and DB-damping are also included from the comparative results reported by \citet{dong2022}. As in the Fabre comparison, these literature results are included to place the performance of the present target-$\omega$ model in the context of existing interfacial turbulence-damping treatments. For the mean streamwise velocity, figure~\ref{fig:staebler-results}($a$) shows that the present model follows the experimental profile closely throughout the gas layer, with particularly good agreement in the vicinity of the gas--liquid interface. By contrast, the no-damping reference result \textcolor{black}{reported by \citet{dong2022}} exhibits a strongly distorted velocity distribution, with the gas velocity substantially underpredicted near the interface and overpredicted farther into the gas layer. The PL- and DB-damping models appreciably improve the prediction relative to the no-damping case, but larger discrepancies remain in the immediate interfacial region than for the present model.

The corresponding turbulent kinetic energy profiles are shown in figure~\ref{fig:staebler-results}($b$). The present model reproduces the measured profile reasonably well over most of the gas layer. The main discrepancy occurs in the near-interface region. At the closest available experimental location on the gas side, the predicted turbulent kinetic energy is lower than the measured value by approximately $34\%$. A larger value of $k$ is predicted in the first pure-gas cell adjacent to the interface, where no experimental measurement is available; hence, the local near-interface behaviour cannot be assessed directly from the available data. A slight overprediction is also observed near the upper wall. The no-damping prediction of $k$ is not included in figure~\ref{fig:staebler-results}($b$) because its substantially larger magnitude would obscure the differences among the damping models; the complete profile is reported in figure~6($b$) of \citet{dong2022}. Among the reference damping treatments, the PL-damping result underpredicts $k$ near the interface, whereas the DB-damping result substantially overpredicts it. The present target-$\omega$ formulation therefore provides a more balanced prediction of the measured turbulent kinetic energy distribution.

Taken together, the velocity and turbulent kinetic energy comparisons show that the treatment of interfacial turbulence has a substantial influence on the gas-side solution in this counter-current two-phase configuration. The present formulation reproduces the mean velocity closely and gives a reasonable prediction of the turbulent kinetic energy over most of the gas layer.

\textcolor{black}{A further assessment of the interfacial turbulence-damping treatment is provided by the predicted liquid depth at MP1, which is not prescribed but develops as part of the coupled two-phase solution. To examine the influence of interfacial damping directly, an additional calculation was performed without the interfacial damping source term, using the same numerical configuration as the present target-$\omega$ model. The predicted mean liquid depths are compared with the experimental measurement and the reference results of \citet{hohne2018} in table~\ref{tab:staebler-depth}.}

\textcolor{black}{The experimental liquid depth at MP1 is $10.24\,\mathrm{mm}$ \citep{staebler2007}. The new no-damping calculation predicts $23.04\,\mathrm{mm}$, corresponding to an overprediction of $125.0\%$. In contrast, the present target-$\omega$ model predicts $10.08\,\mathrm{mm}$, with a relative error of approximately $1.6\%$. A similarly pronounced overprediction in the absence of interfacial damping was reported by \citet{hohne2018}, who obtained an MP1 liquid depth of $25.9\,\mathrm{mm}$ without damping and $10.5$--$10.6\,\mathrm{mm}$ with Egorov-type turbulence damping. The present no-damping calculation therefore confirms that omitting interfacial turbulence damping leads to a substantial overprediction of the liquid depth in this counter-current configuration.}

\begin{table*}[!t]
\centering
\caption{\textcolor{black}{Experimental and predicted mean liquid depths at MP1 for St\"abler case~\q{3}. Relative errors are calculated with respect to the experimental value of $10.24\,\mathrm{mm}$.}}\vspace{0.15cm}
\label{tab:staebler-depth}
\begingroup
\renewcommand{\arraystretch}{1.25}
\begin{tabular}{lcc}
\hline
Case & Liquid depth (mm) & Relative error (\%) \\
\hline
St\"abler (2007), experiment & 10.24 & -- \\
Present study, no damping & 23.04 & 125.0 \\
Present study, target-$\omega$ damping & 10.08 & 1.6 \\
\shortstack[l]{H\"ohne and Porombka (2018),\\no damping} & 25.9 & 152.9 \\
\shortstack[l]{H\"ohne and Porombka (2018),\\with Egorov-type damping} & 10.5--10.6 & 2.5--3.5 \\
\hline
\end{tabular}
\endgroup
\end{table*}

\textcolor{black}{The calculations of \citet{hohne2018} employed an Egorov-type damping treatment within an Euler--Euler two-fluid framework, in which an empirical damping coefficient must be specified. By contrast, the present target-$\omega$ formulation determines its gas-side target value from the locally resolved flow through the wall-law inversion, using fixed constants without case-specific adjustment. Its response to different interface velocities and near-interface grid resolutions has been assessed against the canonical DNS cases, while the Fabre and St\"abler configurations provide additional experimental assessments without retuning.
Given that the two studies employ different multiphase frameworks and numerical configurations, their liquid-depth predictions do not constitute a direct one-to-one comparison of the damping closures. Nevertheless, the present calculations demonstrate that the target-$\omega$ treatment substantially improves the liquid-depth prediction relative to the no-damping calculation performed within the same numerical framework.}

\begin{figure*}[!t]
\centering
\includegraphics[width=0.9\linewidth]{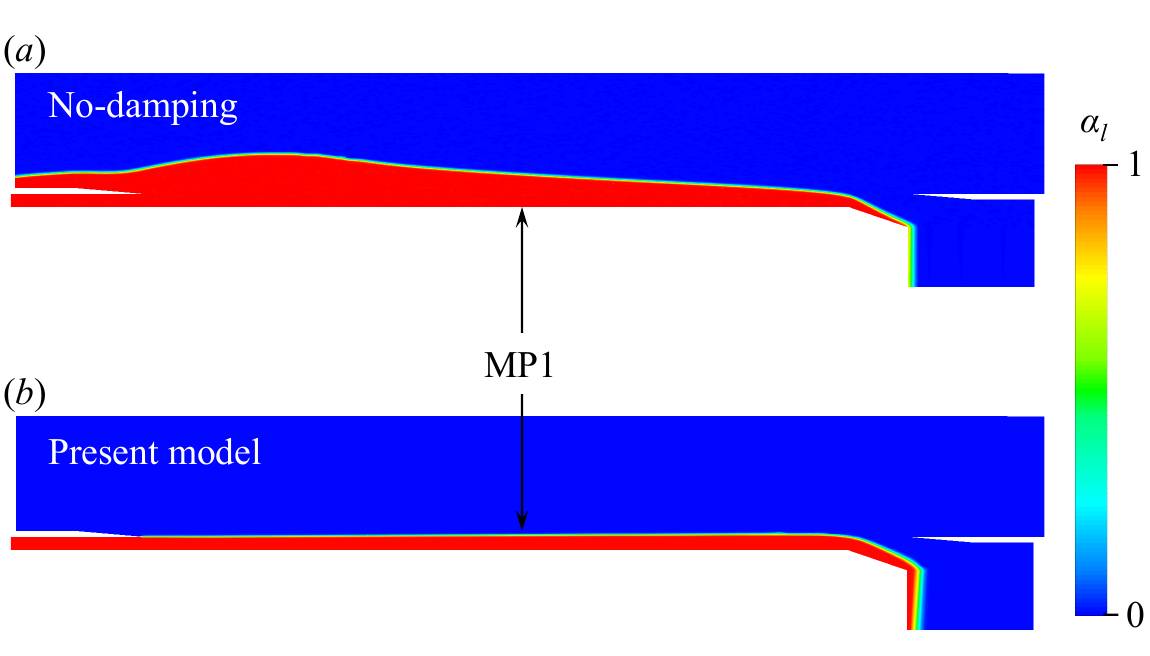}
\caption{\textcolor{black}{Comparison of the predicted liquid-phase volume-fraction fields for St\"abler case~\q{3}: ($a$) without interfacial turbulence damping and ($b$) with the present target-$\omega$ formulation. The arrows indicate the location of MP1. Both panels use the same liquid volume-fraction scale, $0\leq\alpha_l\leq1$.}}
\label{fig:staebler-interface}
\end{figure*}

\textcolor{black}{The influence of interfacial turbulence damping on the predicted interface shape is further illustrated in figure~\ref{fig:staebler-interface}. Without damping, the liquid layer exhibits a pronounced elevation upstream of MP1 and a substantially deformed interface. In contrast, the present target-$\omega$ formulation produces a much flatter interface, with the predicted liquid depth at MP1 in close agreement with the experimental measurement. The marked difference between the two calculations demonstrates that the interfacial turbulence treatment affects not only the gas-side velocity and turbulence fields but also the overall predicted phase distribution.}

The St\"abler assessment therefore extends the evaluation beyond configurations in which the interface geometry is prescribed \emph{a priori}. Within the present scope of resolved large gas--liquid interfaces, the results indicate that the target-$\omega$ formulation can be transferred to a counter-current two-phase configuration with a spatially varying interface without case-specific adjustment. Regimes dominated by breakup, entrainment or dispersed-interface turbulence modulation remain outside the scope of the present model.
\end{revision}

\section{Conclusions and outlook}
\label{sec:conclusions}

This study has addressed the modelling of turbulence damping near resolved large gas--liquid interfaces in Reynolds-averaged simulations. In standard eddy-viscosity closures, strong mean velocity gradients near a density-discontinuous interface may lead to excessive turbulence production and, consequently, to an overprediction of the turbulent viscosity in the interfacial region. Existing damping treatments, such as Egorov-type and related source-term formulations \citep{egorov2004,frederix2018,fan2019,varallo2024}, mitigate this issue by imposing a wall-like enhancement of turbulence dissipation near the interface. However, their effective damping strength depends on empirical parameters and, in some formulations, on the near-interface grid resolution.

To address these limitations, a target-$\omega$ interfacial damping formulation has been proposed using the low-Reynolds-number Wilcox-type $k$--$\omega$ model as the underlying RANS closure. The key idea is to construct a gas-side target value of the specific dissipation rate from an effective interfacial wall coordinate, rather than prescribing the damping strength directly through an Egorov-type empirical prefactor. The model is localised in the gas-side interfacial region and acts through an additional source term in the one-field $\omega$ equation. \textcolor{black}{The constants entering the target-$\omega$ relation are fixed \emph{a priori}, using values adopted from the underlying near-wall treatment and turbulence closure. No constant in the target-$\omega$ relation is fitted to the DNS or experimental data considered in the present study.}

\textcolor{black}{The formulation was first assessed against three canonical DNS cases at $Re_\tau\simeq180$, corresponding to $u_I^+=0$, $0.95$ and $1.9$. Over this range of interface velocities, the model reproduced the principal features of the mean velocity, turbulent kinetic energy and Reynolds shear-stress distributions. The predictions obtained on the finer and intermediate grids (i.e. for $y_{\mathrm{grid}}^+\leq3$, with $y_{\mathrm{grid}}^+$ defined using the interface-normal cell size $\Delta_\perp$) remained comparatively insensitive to the near-interface resolution, while a more noticeable grid dependence appeared on the coarsest mesh ($y_{\mathrm{grid}}^+=6$), particularly in the near-boundary peaks of turbulent kinetic energy. The present formulation therefore reduces, but does not eliminate, sensitivity to the near-interface grid resolution. By comparison, the Egorov predictions exhibited a substantially stronger dependence on both the prescribed damping factor $B$ and the near-interface grid spacing. These results indicate that the target-$\omega$ formulation provides a more consistent control of the gas-side interfacial turbulence over the canonical conditions considered here without case-specific adjustment of a damping coefficient.}

\textcolor{black}{The model was then assessed against two stratified air--water experiments. For the co-current Fabre case ``250'' \citep{fabre1987experimental}, which has a higher friction Reynolds number of approximately $Re_\tau\simeq500$, the present formulation reproduced the measured gas-side velocity and Reynolds shear-stress distributions well and gave a reasonable prediction of the turbulent kinetic energy. Its performance was comparable to the calibrated phenomenological length-scale damping reference reported by \citet{dong2022}, without any Fabre-specific adjustment of the target-$\omega$ formulation. A further assessment was performed against the counter-current St\"abler case \citep{staebler2007}, for which the liquid level is not prescribed but develops as part of the coupled two-phase solution. The present model reproduced the measured gas-side velocity profile closely and gave a reasonable turbulent kinetic energy distribution over most of the gas layer. The predicted mean liquid depth at MP1 was $10.08\,\mathrm{mm}$, compared with the experimental value of $10.24\,\mathrm{mm}$, corresponding to a difference of approximately $1.6\%$. This latter case indicates that the formulation can also be applied in a coupled two-phase configuration in which the interface position is part of the solution rather than prescribed \emph{a priori}.}

Although the present assessments show favourable behaviour within the cases considered, several aspects remain open for future work. First, the formulation should be tested over a wider parameter range, including Reynolds number, interface velocity, and density and viscosity ratios. \textcolor{black}{The present assessments remain focused on resolved large interfaces for which a wall-like constraint on the gas-side turbulence is a plausible leading-order approximation. Although the St\"abler case extends the evaluation to a coupled counter-current configuration with a spatially varying interface, more strongly wavy or highly deformed interfaces may involve additional turbulence--interface interactions that are not represented explicitly in the present formulation \citep{fulgosi2003,berthelsen2005,benz2015,dong2022}.} Another important extension concerns regimes in which the liquid-side turbulence is no longer weak. In such cases, the interface may no longer behave as a purely wall-like constraint for the lighter phase, and the relative damping strength on the two sides of the interface may need to be treated asymmetrically \citep{liu2009,frederix2018,dong2022}.

A further extension would be to incorporate the target-$\omega$ idea into an Euler--Euler two-fluid framework \citep{drew1983mathematical}. The present model has been developed in a one-field interface-capturing setting, where the large-scale interface is explicitly represented. However, many gas--liquid flows of practical interest are intrinsically multi-scale: they may contain large-scale continuous interfaces together with much smaller dispersed bubbles or droplets, with length scales differing by orders of magnitude. Euler--Euler two-fluid or multifield models provide a useful framework for such flows because they solve separate phase-balance equations and represent interfacial momentum transfer, coalescence, breakup and turbulence modulation through closure relations. This direction is closely related to recent multi-scale and morphology-adaptive approaches, in which large interfaces and dispersed structures are treated within a unified framework \citep{mer2018emptying,mathur2019,frederix2021,meller2023,garcia2025}. Incorporating a physically constrained interfacial turbulence-damping model into such approaches would therefore be a natural step towards simulations of mixed-regime flows, where large-scale interfaces, entrainment, detrainment and dispersed-phase dynamics coexist.

Finally, future work may couple the present damping formulation with heat and mass transfer models. In the underlying physical processes, interfacial heat and mass transfer are closely linked to interfacial shear, turbulence, wave breaking and the formation of droplets or bubbles \citep{deike2022mass}. In RANS scalar-transport modelling, these effects are represented through closure quantities such as the turbulent viscosity, turbulent diffusivity and near-interface turbulence time scale. An improved prediction of near-interface turbulence damping may therefore affect the accuracy of modelled scalar-transfer predictions. This would be particularly relevant for evaporating, condensing or surfactant-laden gas--liquid flows, where turbulence, interfacial deformation and scalar transport are closely coupled.

\section*{Acknowledgements}

This work was supported by the National Key Research and Development Program of China under Grant No.~2025YFE0220100. P.~Shi acknowledges support from the Fonds de la Recherche Scientifique -- FNRS under Grant No.~40031630, during the period in which the present manuscript was prepared.

\section*{Declaration of generative AI and AI-assisted technologies in the manuscript preparation process}

During the preparation of this work, the authors used ChatGPT (OpenAI) to assist with language editing. The tool was not used to generate the numerical results, figures, simulations or scientific conclusions. The authors reviewed and edited the output as needed and take full responsibility for the content of the published article.

\appendix
\section{Existing interfacial turbulence-damping treatments}
\label{app:damping_models}

This appendix summarises representative interfacial turbulence-damping treatments used in Reynolds-averaged simulations of large gas--liquid interfaces. The purpose is not to provide an exhaustive review, but to clarify the modelling ideas behind the reference formulations discussed in the main text and to highlight their relation to the present target-$\omega$ model.

\subsection*{Wall-like damping in low-Reynolds-number $k$--$\varepsilon$ models}

A direct way of suppressing excessive turbulence near a smooth gas--liquid interface is to treat the interface as an additional damping surface in a low-Reynolds-number $k$--$\varepsilon$ model. This approach was used by \citet{berthelsen2007} for smooth stratified two-phase pipe flow and was later adopted by \citet{duan2014} for stratified gas--liquid pipe flow with heat transfer.

In such models, the turbulent viscosity is commonly written as
\begin{equation}
\label{eq:mut_lre_ke}
\mu_t=\rho C_\mu f_\mu\frac{k^2}{\varepsilon},
\end{equation}
where $f_\mu$ is a near-boundary damping function. In single-phase wall-bounded turbulence, this function is evaluated using the distance to the nearest solid wall. In the interfacial extension, the relevant distance is replaced by the shortest distance to either the wall or the gas--liquid interface,
\begin{equation}
\label{eq:nearest_surface_distance}
d=\min(d_{\mathrm{wall}},d_I).
\end{equation}
The turbulent viscosity is therefore damped not only near solid walls, but also near the smooth interface.

This class of treatment is physically motivated when the interface behaves as a nearly flat, shear-constraining boundary. However, it also implies that the interface is treated in a wall-like manner for the turbulence field. This assumption becomes less straightforward when the interface is deformable, wavy, or when both phases are turbulent. Moreover, this type of damping is closely tied to the structure of low-Reynolds-number $k$--$\varepsilon$ models and is not directly transferable to all turbulence closures.

\subsection*{Free-surface dissipation condition in a $k$--$\omega$ model}

\citet{kamath2019} proposed a free-surface damping treatment in a two-phase $k$--$\omega$ RANS model. Instead of adding an explicit source term to the $\omega$ equation, they prescribed a near-surface value of the specific dissipation rate. Their formulation may be written as
\begin{equation}
\label{eq:kamath_omega_s}
\omega_s=C_\mu^{-1/4}\frac{\sqrt{k}}{\kappa}
\left(
\frac{1}{y_0}+\frac{1}{y^*}
\right),
\end{equation}
where $y_0$ is a virtual origin for the turbulent length scale and $y^*$ is the distance from the nearest wall. The correction is localised near the interface through a regularised Dirac delta function based on the level-set field.

The essential idea is to constrain the turbulence length scale near the free surface through the imposed value of $\omega_s$. Since the turbulent viscosity in a $k$--$\omega$ model scales with $k/\omega$, increasing $\omega$ near the interface reduces the local turbulent viscosity. This mechanism is similar in spirit to source-term damping models, although the numerical implementation differs because the correction is introduced through a near-surface value of $\omega$ rather than through an additional production term in the $\omega$ equation.

\subsection*{Egorov-type source term in the $\omega$ equation}

\citet{egorov2004} introduced one of the most widely used interfacial turbulence-damping treatments for $k$--$\omega$-type models. The model adds a positive source term to the $\omega$ equation in the interfacial region. By increasing $\omega$, it reduces the turbulent viscosity through the usual $k/\omega$ dependence. The full source-term expression and the definition of its parameters have already been given in Eq.~\eqref{eq:egorov_intro} and are therefore not repeated here.

The Egorov formulation is based on the near-wall behaviour of the specific dissipation rate in a $k$--$\omega$ model. Close to a wall, $\omega$ scales as $\nu/y^2$, where $y$ is the wall-normal distance. In the interfacial source-term formulation, this wall-normal distance is effectively replaced by a local interface-normal grid-size measure, and the imposed value of $\omega$ is multiplied by an empirical damping factor $B$. The model is therefore straightforward to implement in interface-capturing or two-fluid calculations, because the source term can be activated by the local interfacial area density.

The main limitation of the original Egorov model is that its effective damping strength is not controlled by a purely physical length scale. Instead, it depends on the empirical factor $B$ and on the local interfacial grid size. Increasing $B$ strengthens the damping by increasing the imposed value of $\omega$, whereas coarsening the grid can weaken the imposed damping through the grid-size dependence of the interfacial source. As discussed in the main text, this explains why a wide range of $B$ values has been used in the literature and why the model may require case- and grid-dependent calibration.

\subsection*{Length-scale reformulation by Frederix et al.}

\citet{frederix2018} analysed the Egorov-type damping concept in a two-fluid Euler--Euler framework and highlighted its dependence on both the empirical damping factor and the near-interface grid size. To reinterpret this dependence in terms of a physically meaningful length scale, they introduced an interfacial damping length $\delta$ defined by
\begin{equation}
\label{eq:frederix_delta}
\delta^2=\frac{\Delta x^2}{6B},
\end{equation}
where $\Delta x$ is the grid size entering the original Egorov-type formulation. With this definition, changing $B$ at fixed grid resolution is equivalent to changing the imposed interfacial damping length.

The corresponding interfacial values of the turbulence variables can be expressed as
\begin{equation}
\label{eq:frederix_omega_epsilon_delta}
\omega_i^{\mathrm{int},*}
=
\frac{\nu_i}{\beta^*\delta^2},
\qquad
\varepsilon_i^{\mathrm{int},*}
=
\frac{k_i\nu_i}{\delta^2},
\end{equation}
where the subscript $i$ denotes the phase. In this interpretation, a smaller $\delta$ corresponds to stronger interfacial dissipation, whereas a larger $\delta$ weakens the damping.

This reformulation is useful because it shifts the calibration from the numerical parameter pair $(\Delta x,B)$ to an effective phenomenological length scale. However, the value of $\delta$ still has to be prescribed or calibrated. In the terminology of \citet{dong2022}, this approach is referred to as phenomenological length-scale damping, or PL-damping. In the Fabre comparison presented in \S~\ref{sec:results_fabre}, the PL-damping curve digitised from \citet{dong2022} should therefore be interpreted as a calibrated Frederix-type reference result.

\subsection*{Fan--Anglart modification of the Egorov model}

\citet{fan2019} proposed a modification of the Egorov model for annular gas--liquid flow. Their work addressed two important limitations of the original source-term formulation: its dependence on the mesh size and its symmetric treatment of the two phases.

To reduce the sensitivity to tangential mesh dimensions, they introduced an interface-normal length scale based on the cell volume and on the projected interfacial area,
\begin{equation}
\label{eq:fan_yp}
y_p=\frac{V}{S_p},
\qquad
S_p=\frac{1}{2}\sum_f |\boldsymbol{n}\cdot\boldsymbol{S}_f|,
\end{equation}
where $V$ is the cell volume, $\boldsymbol{n}$ is the interface normal and $\boldsymbol{S}_f$ is the face-area vector. This definition is intended to provide a local interface-normal measure that is less directly affected by stretching of the mesh in directions tangential to the interface.

\citet{fan2019} also proposed an asymmetric damping treatment in which the source term is weighted differently on the gas and liquid sides. This is important because the turbulence response to an interface is generally not symmetric when the density and viscosity ratios are large. Their formulation therefore retains the practical source-term structure of the Egorov model, but introduces a more flexible localisation and a phase-dependent damping strength.

\subsection*{Egorov-type correction in the $\varepsilon$ equation}

Although Egorov-type damping is most commonly associated with $k$--$\omega$ models, analogous ideas have also been introduced in $k$--$\varepsilon$ closures. \citet{varallo2024} proposed an interfacial damping correction for an RNG $k$--$\varepsilon$ model by adding a wall-like source term to the $\varepsilon$ equation.

The purpose of this correction is analogous to that of the Egorov source term in the $\omega$ equation: the model increases the local dissipation rate in the interfacial region and thereby reduces excessive turbulent viscosity generated by large interfacial velocity gradients. The formulation is activated through the interfacial area density and includes an empirical damping factor. Therefore, although it is formulated in terms of $\varepsilon$ rather than $\omega$, it retains the same general modelling ingredients as Egorov-type damping: interfacial localisation, wall-like dissipation enhancement and empirical control of the damping strength.

\subsection*{Density-based damping in two-fluid turbulence models}

\citet{dong2022} proposed a density-based damping model for two-fluid RANS simulations. Their formulation was motivated by the observation that interfacial damping should depend on the local phase distribution and on the density contrast between the two phases, rather than solely on an empirical damping factor or an imposed phenomenological length scale.

In their model, the damping strength is controlled by a coefficient of the form
\begin{equation}
\label{eq:dong_cd}
C_{d,i}=
\frac{(\rho_i-\rho_j)\alpha_i-\rho_m}{\rho_m},
\end{equation}
where $\alpha_i$ and $\rho_i$ are the volume fraction and density of phase $i$, $j$ denotes the other phase and $\rho_m$ is the mixture density. The corresponding damping contributions are introduced into the per-phase turbulence transport equations of the two-fluid model.

A key feature of the density-based formulation is that it naturally produces different damping strengths in the two phases. In gas--liquid flows with large density contrast, the model tends to damp the turbulence more strongly in the lighter phase, while applying weaker damping in the denser phase. This behaviour is consistent with the expectation that the interface acts more like a constraint for the less dense gas phase than for the liquid phase. Since the damping coefficient is constructed from local phase and density information, the model avoids prescribing an Egorov-type empirical prefactor $B$. However, it is formulated within a two-fluid framework and is therefore not directly identical to one-field interface-capturing damping models such as the present target-$\omega$ formulation.

\balance
\bibliographystyle{elsarticle-harv}
\bibliography{references_ijmf_updated}

\end{document}